\documentclass{svcyclop}
\usepackage{latexsym}
\usepackage{graphicx}

\usepackage[T1]{fontenc}
\usepackage[latin9]{inputenc}
\usepackage{lmodern} 
\usepackage{listings}
\usepackage[numbers]{natbib}
\usepackage{authblk}
\usepackage{url}
\usepackage{graphicx}
\usepackage{amsmath,amssymb,amsfonts,graphicx,latexsym}
\usepackage[english]{babel}
\usepackage{verbatim}

\DeclareGraphicsExtensions{.jpg,.pdf,.png}

\newcommand{\ot}{OpenTURNS }
\newcommand{\pyobj}[1]{\texttt{#1}}
\newcommand{\bm}[1]{\mbox{\boldmath $#1$}}
\newcommand{\mb}[1]{\mathbf{#1}}

\newcommand{\R}{\mathbb{R}}     
\newcommand{\N}{\mathbb{N}}         

\newcommand{\EE}{\mathbb{E}}      

\newcommand{\vecX}{(X_1,\ldots, X_d)} 
\newcommand{\vecx}{(x_1,\ldots, x_d)} 

\begin{document}

\title{\ot: An industrial software for uncertainty quantification in simulation}

\author{Micha\"el Baudin, Anne Dutfoy, Bertrand Iooss and Anne-Laure Popelin}

\institute{
EDF R\&D\\
6 quai Watier, 78401 Chatou, France\\
1, avenue du General de Gaulle Clamart, France\\
E-mail: michael.baudin@edf.fr, anne.dutfoy@edf.fr, bertrand.iooss@edf.fr, anne-laure.popelin@edf.fr
}

\maketitle

\section*{Abstract}

The needs to assess robust performances for complex systems and to answer tighter regulatory processes (security, safety, environmental control, and health impacts, etc.) have led to the emergence of a new industrial simulation challenge: to take uncertainties into account  when dealing with complex numerical simulation frameworks. 
Therefore, a generic methodology has emerged from the joint effort of several industrial companies and academic institutions.
EDF R\&D, Airbus Group and Phimeca Engineering  started a collaboration at the beginning of  2005, joined by IMACS in 2014, for the development of an Open Source software platform dedicated to uncertainty propagation by probabilistic methods, named \ot for Open source Treatment of Uncertainty, Risk 'N Statistics. 
\ot addresses the specific industrial challenges attached to uncertainties, which are transparency, genericity,  modularity and multi-accessibility.
This paper focuses on \ot and presents its main features: \ot is an open source software under the LGPL license, that presents itself as a C++ library and a Python TUI, and which works under Linux and Windows environment. All the methodological tools are described  in the different sections of this paper:  uncertainty quantification,  uncertainty propagation,  sensitivity analysis and  metamodeling. A section  also explains the generic wrappers way to link \ot to any external code.
The paper illustrates as much as possible the methodological tools on an educational example that simulates the height of a river and compares it to the height of a dyke that protects industrial facilities.
At last, it gives an overview of the main developments planned for the next few years. \\

{\bf Key words}: \ot, uncertainty, quantification, propagation, estimation, sensitivity, simulation, probability, statistics, random vectors, multivariate distribution, open source, python module, C++ library, transparency, genericity.

\section{Introduction}

The needs to assess robust performances for complex systems and to answer tighter regulatory processes (security, safety, environmental control, and health impacts, etc.) have led to the emergence of a new industrial simulation challenge: to take uncertainties into account  when dealing with complex numerical simulation frameworks. 
Many attempts at treating uncertainty in large industrial applications have involved domain-specific approaches or standards: metrology, reliability, differential-based approaches, variance decomposition, etc.
However, facing the questioning of their certification authorities in an increasing number of different domains, these domain-specific approaches are no more appropriate.
Therefore, a generic methodology has emerged from the joint effort of several industrial companies and academic institutions : \citep{pas14} reviews these past developments. 
The specific industrial challenges attached to the recent uncertainty concerns are:
\begin{itemize}
\item transparency: open consensus that can be understood by outside authorities and experts,
\item genericity: multi-domain issue that involves various actors along the study,
\item modularity: easy integration of innovations from the open source community,
\item multi-accessibility: different levels of use (simple computation, detailed quantitative results, and deep graphical analyses) and different types of end-users (graphical interface, Python interpreter, and C++ sources),
\item industrial computing capabilities: to secure the challenging number of simulations required by uncertainty treatment.
\end{itemize}

As no software was fully answering the challenges mentioned  above, EDF R\&D, Airbus Group and Phimeca Engineering  started a collaboration at the beginning of  2005, joined by EMACS in 2014, for the development of an open Source software platform dedicated to uncertainty propagation by probabilistic methods, named \ot for open source Treatment of Uncertainty, Risk 'N Statistics \citep{dutdut09},\citep{pasdut12}.
\ot is actively supported by its core team of four industrial partners (IMACS joined the consortium in 2014) and its industrial and academic users community that meet through the web site \url{www.openturns.org} and annually during the \ot Users day.
At EDF, \ot is the repository of all scientific developments on this subject, to ensure their dissemination within the several business units of the company. 
The software has also been distributed for several years via the integrating platform Salome \citep{ope06}.

\subsection{Presentation of \ot}

\ot is an open source software under the LGPL license, that presents itself as a C++ library and a Python TUI, and which works under Linux and Windows environment, with the following key features:
\begin{itemize}
\item open source initiative to secure the transparency of the approach, and its openness to ongoing
Research and Development (R\&D)  and expert challenging,
\item generic to the physical or industrial domains for treating of multi-physical problems;
\item structured in a \emph{practitioner-guidance} methodological approach,
\item with advanced industrial computing capabilities, enabling the use of massive distribution
and high performance computing, various engineering environments, large data models etc.,
\item includes the largest variety of qualified algorithms in order to manage uncertainties in several situations,
\item contains complete documentation (Reference Guide, Use Cases Guide, User manual, Examples Guide,  and Developers' Guide).
\end{itemize}
All the methodological tools are described after this introduction in the different sections of this paper:  uncertainty quantification,  uncertainty propagation,  sensitivity analysis and  metamodeling. 
Before the conclusion, a section  also explains the generic wrappers way to link \ot to any external code.

\ot can be downloaded from its dedicated website \emph{www.openturns.org} which offers different pre-compiled packages specific to several Windows and Linux environments. It is also possible to download the source files from the  \emph{SourceForge} server (\emph{www.sourceforge.net})  and to compile them within another environment: the \ot Developer's Guide provides advice to help compiling the source files. At last, \ot has been integrated for more than 5 years in the major Linux distributions (for example debian, ubuntu, redhat and suze).

\subsection{The uncertainty management methodology}

\begin{figure}[!ht]
\begin{center}
\includegraphics[width=13cm]{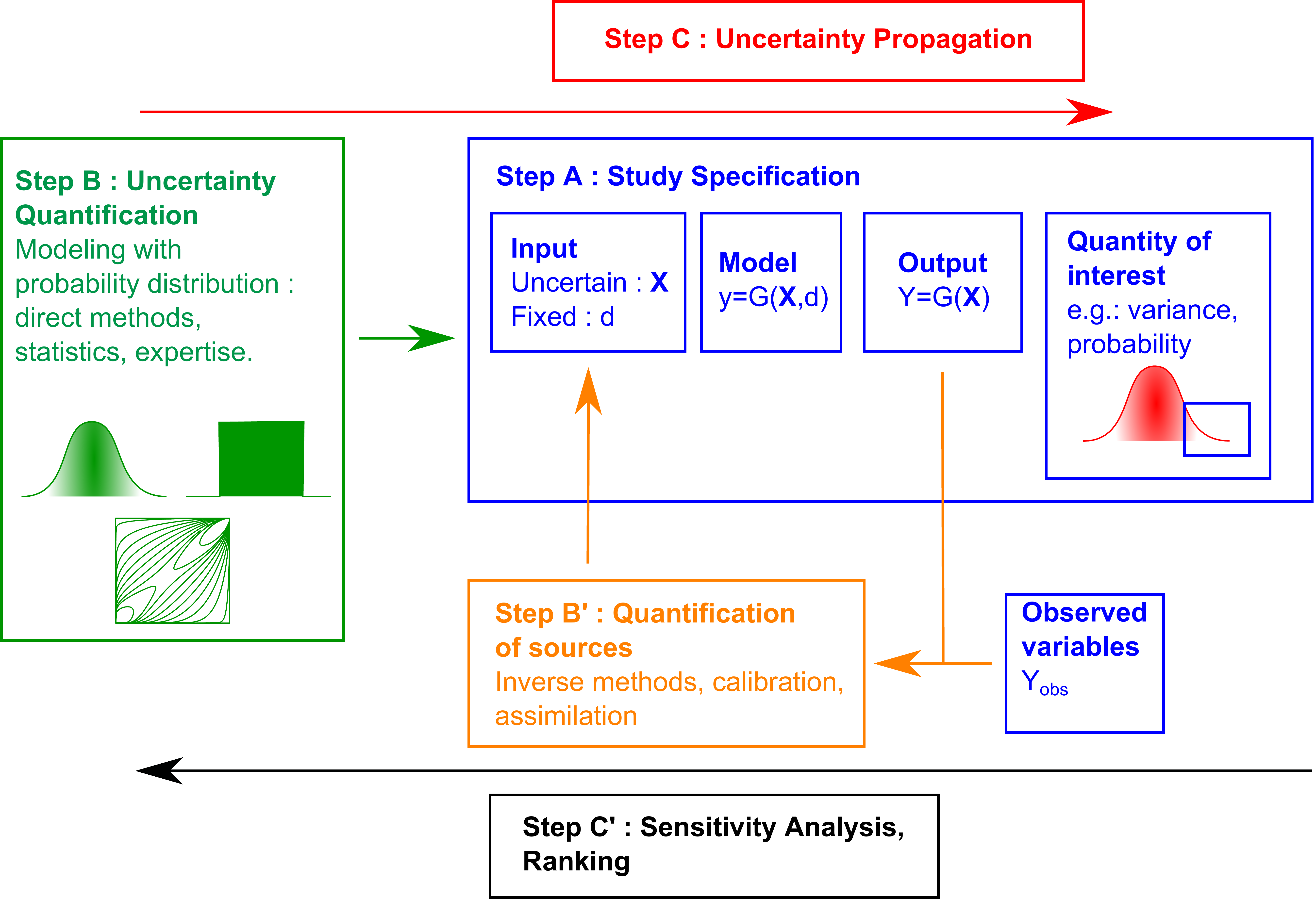}
\caption{The uncertainty management methodology.}\label{fig:methodo}
\end{center}
\end{figure}

The uncertainty management generic methodology \citep{pasdut12} is schematized in Figure \ref{fig:methodo}.
It consists of the following steps:
\begin{itemize}
\item Step A: specify the random inputs $\bm{X}$, the deterministic inputs $\bm{d}$, the model $G$ (analytical, complex computer code or experimental process), the variable of interest (model output) $Y$ and the quantity of interest on the output (central dispersion, its distribution, probability to exceed a threshold, \ldots). 
The fundamental relation writes:
\begin{equation}
Y = G(\bm{X},\bm{d}) = G(\bm{X}),
\end{equation}
with $\bm{X} = (X_1,\ldots,X_d)$.
\item Step B: quantify the sources of uncertainty. This step consists in modeling the joint probability density function (pdf) of the random input vector by direct methods (e.g. statistical fitting, expert judgment) \citep{kurcoo06}. 
\item Step B': quantify the sources of uncertainty by indirect methods using some real observations of the model outputs \citep{tar05}.
The calibration process aims to estimate the values or the pdf of the inputs while the validation process aims to model the bias between the model and the real system.
\item Step C: propagate uncertainties to estimate the quantity of interest. With respect to this quantity, the computational resources and the CPU time cost of a single model run, various methods will be applied: analytical formula, geometrical approximations, Monte Carlo sampling strategies, metamodel-based techniques, \ldots \citep{lema09}, \citep{fanli06}.
\item Step C': analyze the sensitivity of the quantity of interest to the inputs in order to rank uncertainty sources \citep{salcha00}, \citep{ioolem15}.  
\end{itemize}
For each of these steps, \ot offers a large number of different methods whose applicability depend on the specificity of the problem (dimension of inputs, model complexity, CPU time cost for a model run, quantity of interest, etc.).

\subsection{Main originality of \ot}
 
\ot is innovative in several aspects. Its input data model is based on
the multivariate cumulative distribution function (CDF). This enables the usual
sampling approach, as would be appropriate for statistical manipulation of large
data sets, but also facilitates analytical approaches.
Distributions are classified (continuous, discrete, elliptic, etc.) in order to take the best benefit of their properties in algorithms.  If possible, the exact final cumulative density function is determined (thanks to characteristic functions implemented for each distribution, the Poisson summation formula, the Cauchy integral formula, etc.).

\ot explicitly models the dependence with copulas, using the Sklar theorem. Furthermore, different sophisticated analytical treatments may be explored: aggregation of copulas, composition of functions from $R^n$ into $R^d$, extraction of copula and marginals from any distribution.

\ot defines a domain specific oriented object language for probability modelling and uncertainty management.
This way, the objects correspond to mathematical concepts and their inter-relations map the relations between these mathematical concepts. Each object proposes sophisticated treatments in a very simple interface.

\ot implements up-to-date and efficient sampling algorithms (Mersenne-Twister algorithm, Ziggurat method, the Sequential Rejection Method, etc.). Exact Kolmogorov statistics are evaluated with the Marsaglia Method and the Non Central Student and Non Central $\chi^2$ distribution with the Benton \& Krishnamoorthy method.

\ot is the repository of recent results of PhD research carried out at EDF R\&D: for instance the sparse Polynomial Chaos Expansion method based on the LARS method \citep{blasud11}, the Adaptive Directional Stratification method \citep{mungar11} which is an accelerated Monte Carlo sampling technique, and the maximum entropy order-statistics copulas \citep{lebdut14}.

\subsection{The flooding model}

\begin{figure}[!ht]
\begin{center}
    \includegraphics[width=9cm]{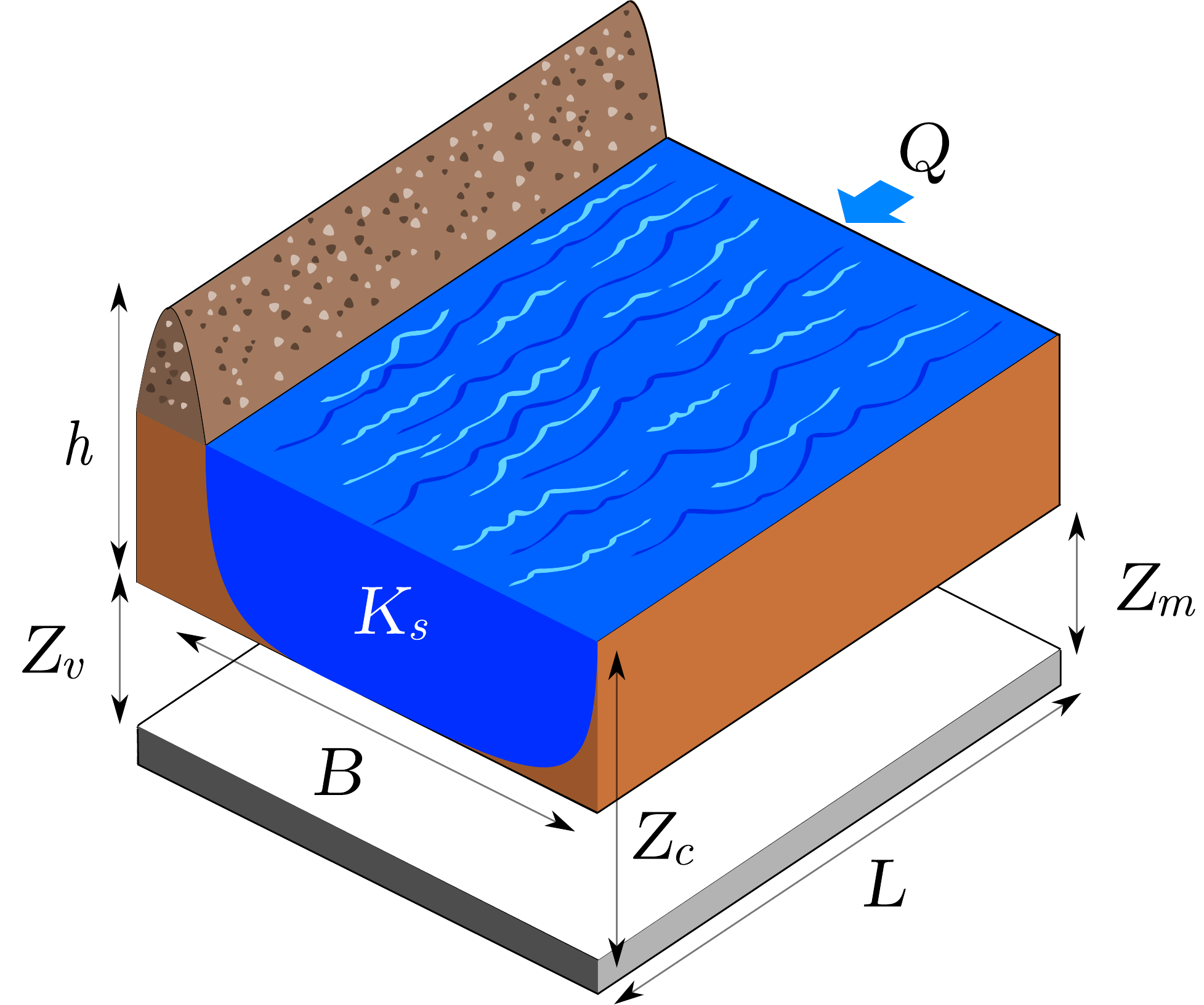} 
\end{center}
\caption{The flood example: simplified model of a river.}\label{fig:crues}
\end{figure}

Throughout this paper,  the discussion is illustrated with a simple application model that simulates the height of a river and compares it to the height of a dyke that protects industrial facilities as illustrated in Figure \ref{fig:crues}. When the river height exceeds that of the dyke, flooding occurs. This academic model is used as a pedagogical example in \citep{ioolem15}.
The model is based on a crude simplification of the 1D hydro-dynamical equations of SaintVenant under the assumptions of uniform and constant flowrate and large rectangular sections. It consists of an equation that involves the characteristics of the river stretch:
\begin{equation}\label{eq:cruesS}
 \displaystyle H = \left(\frac{Q}{B K_s \sqrt{\frac{Z_m-Z_v}{L} }} \right)^{0.6},
\end{equation}
where  the output variable $H$ is the maximal annual height of the river, $B$ is the river width and $L$ is the length of the river stretch. 
The four random input variables $Q$, $K_s$, $Z_v$ and $Z_m$ are defined in Table \ref{tab:factors} with their  probability distribution.
The randomness of these variables is due to their spatio-temporal variability, our ignorance of their true value or some inaccuracies of their estimation. 

\begin{table}[!ht]
  \begin{center}
   \begin{tabular}{lccc}
Input & Description & Unit & Probability distribution \\
   \hline
 $Q$ & Maximal annual flowrate & m$^3$/s & Gumbel ${\mathcal G}(1.8e^{-3}, 1014)$ \\
 $K_s$ & Strickler coefficient & - & Normal ${\mathcal N}(30, 7.5)$  \\
 $Z_v$ & River downstream level & m & Triangular  ${\mathcal T}(47.6, 50.5, 52.4)$ \\
 $Z_m$ & River upstream level  & m  & Triangular  ${\mathcal T}(52.5, 54.9, 57.7)$  \\
 \hline
    \end{tabular}
    \caption{Input variables of the flood model and their probability distributions.}
    \label{tab:factors}
  \end{center}
\end{table}

\section{Uncertainty quantification}\label{UncertaintyQuantification}


\subsection{Modelling of a random vector}

\begin{figure}[!ht]
\begin{center}
\includegraphics[width=7.5cm]{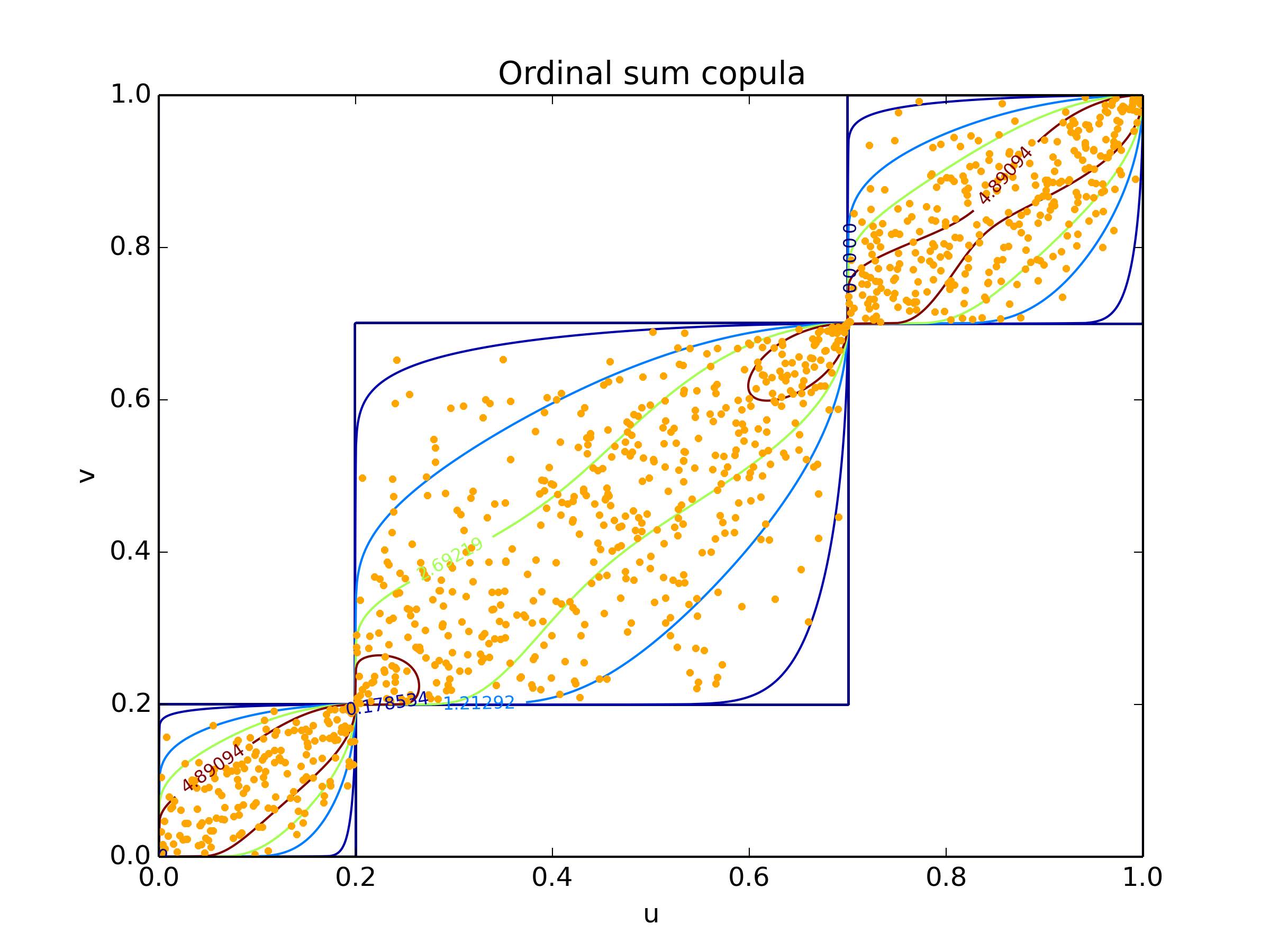}
\caption{An example of  maximum entropy copula which  almost surely satisfies the ordering constraint: $X_1 \leq X_2$.}
\label{distOS}
\end{center}
\end{figure}

\ot  implements more than 40 parametric distributions which are continuous (more than 30 families) and discrete (more than 10 families), with several sets of parameters for each one. Some are multivariate, such as the Student distribution or the Normal one. \\
Moreover, \ot enables the building of  a wide variety of multivariate distributions thanks to the combination of the univariate margins and a dependence structure, the copula, according to the Sklar theorem: $F\vecx = C(F_1(x_1), \dots, F_d(x_d))$ where $F_i$ is the CDF of the margin $X_i$ and $C: [0,1]^d \rightarrow [0,1]$ the copula.\\
 \ot proposes more than 10 parametric families of copula: Clayton, Frank, Gumbel, Farlie-Morgenstein, etc. These copula can be aggregated to build the copula of a random vector whose components are dependent by blocks. Using the inverse relation of the Sklar theorem, \ot can extract the copula of any multivariate distribution, whatever the way it has been set up: for example, from a multivariate distribution estimated from a sample with the kernel smoothing technique.\\
All the distributions can be truncated in their lower and/or upper area.
In addition to these models, \ot proposes other specific constructions. Among them, note the random vector which writes as a linear combination of a finite set of independent variables: $\bm{X} =  a_0 + a_1 \bm{X}_1 + \dots a_N \bm{X}_N$ thanks to the python command, written for $N=2$ with explicit notations:
{\small
\lstset{language=Python}
\begin{lstlisting}
>>>myX= RandomMixture([distX1,distX2], [a1, a2], a0)
\end{lstlisting}
}
 In that case, the distribution of $X$ is \emph{exactly} determined, using the characteristic functions of the $X_i$ distributions and the Poisson summation formula.\\
\ot also easily models the random vector whose probability density function (pdf) is a linear combination of a finite set of independent pdf: $f_{\bm{X}} =  a_1 f_{\bm{X}_1} + \dots a_N f_{\bm{X}_N}$ thanks to the python command, with the same notations as previously (the weights are automatically normalized):
{\small
\lstset{language=Python}
\begin{lstlisting}
>>>mypdfX= Mixture([distX1,distX2], [a1, a2])
\end{lstlisting}
}
Moreover, \ot implements a random vector that writes as the random sum of univariate independent and identically distributed variables, this randomness being distributed according to a Poisson distribution: $\bm{X} =  \sum_{i=1}^N \bm{X}_i, \quad N\sim \mathcal{P}(\lambda)$,  thanks to the python command:
{\small
\lstset{language=Python}
\begin{lstlisting}
>>>d= CompoundDistribution(lambda, distX)
\end{lstlisting}
}
where all the variables $X_i$ are identically distributed according to $distX$. In that case, the distribution of $X$ is \emph{exactly} determined, using the characteristic functions of the $X_i$ distributions and the Poisson summation formula.\\
In the univariate case, \ot exactly determines the pushforward distribution $\mathcal{D}$ of any distribution $\mathcal{D}_0$ through the function $f:\R \rightarrow \R$, thanks to the python command (with straight notations):
{\small
\lstset{language=Python}
\begin{lstlisting}
>>>d= CompositeDistribution(f,d0)
\end{lstlisting}
}
Finally,  \ot enables the modeling of a random vector $\vecX$ which almost surely verifies the constraint $\bm{X} = X_1 \leq \dots \leq X_d$ , proposing a copula adapted to the ordering constraint \citep{Fischer2015}. \ot verifies the compatibility of the margins $F_i$ with respect to the ordering constraint and then builds the associated distribution, thanks to the python command, written in dimension 2:
{\small
\lstset{language=Python}
\begin{lstlisting}
>>>d=MaximumEntropyOrderStatisticsDistribution([distX1, distX2])
\end{lstlisting}
}
Figure \ref{distOS} illustrates the copula of such a distribution, built as the ordinal sum of some maximum entropy order statistics copulae.

The \ot python script to model the input random vector of the tutorial presented previously is as follows:
{\small
\lstset{language=Python}
\begin{lstlisting}
#Margin distributions:
>>>dist_Q = Gumbel(1.8e-3, 1014)
>>>dist_Q = TruncatedDistribution(dist_Q,0.0, TruncatedDistribution.LOWER)
>>>dist_K = Normal(30.0, 7.5)
>>>dist_K = TruncatedDistribution(dist_K,0., TruncatedDistribution.LOWER)
>>>dist_Zv = Triangular(47.6,50.5,52.4)
>>>dist_Zm = Triangular(52.5,54.9,57.7)
# Copula in dimension 4 for (Q,K,Zv,Zm)
>>>R=CorrelationMatrix(2)
>>>R[0,1]=0.7
>>>copula = ComposedCopula([IndependentCopula(2),NormalCopula(R) ])
# Final distribution for (Q,K,Zv,Zm)
>>>distInput=ComposedDistribution([loi_Q, loi_K, loi_Zv, loi_Zm], copula)
# Final random vector (Q,K,Zv,Zm)
>>>inputVector=RandomVector(distInput)
\end{lstlisting}
}
Note that \ot can truncate any distribution to a lower, an upper bound or a given interval. Furthermore, a normal copula models the dependence  between the variables $Z_v$ and $Z_m$, with a correlation of 0.7. The variables $(Q,K)$ are independent. Both blocks $(Q,K)$ and $(Z_v,Z_m)$ are independent.

\subsection{Stochastic processes}

\begin{figure}[!ht]
	  \begin{minipage}{7.5cm}
	    \begin{center}
	      \includegraphics[width=7.5cm]{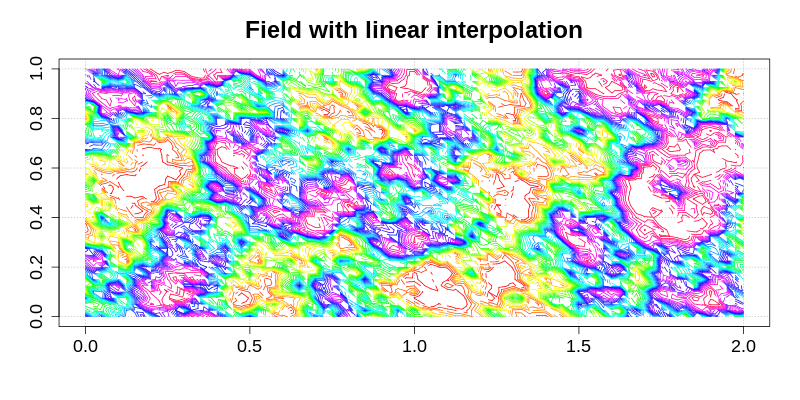}
	      \caption{A Normal bivariate white noise.}
	      \label{whiteNoise}
	    \end{center}
	  \end{minipage}
	  \hfill
	  \begin{minipage}{7.5cm}
	    \begin{center}
	      \includegraphics[width=7.5cm]{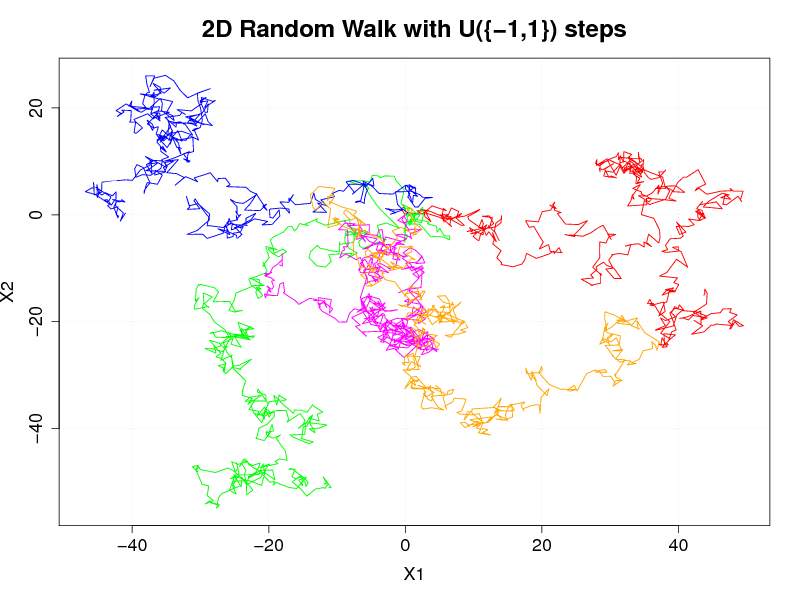}
	      \caption{A Normal bivariate random walk.}
	      \label{RandomWalk}
	    \end{center}
	  \end{minipage}
	\end{figure}

\ot implements some multivariate random fields $\bm{X}: \Omega \times \mathcal{D} \rightarrow \R^{d}$ where $\mathcal{D} \in \R^{s}$ is discretized on a mesh.
The User can easily build and simulate a random walk, a white noise as illustrated in Figures \ref{whiteNoise} and \ref{RandomWalk}. The python commands write:
{\small
\lstset{language=Python}
\begin{lstlisting}
>>>myWN = WhiteNoise(myDist, myMesh)
>>>myRW = RandomWalk(myOrigin, myDist, myTimeGrid)
\end{lstlisting}
}

\begin{figure}[!ht]
\begin{center}
\includegraphics[width=7.5cm]{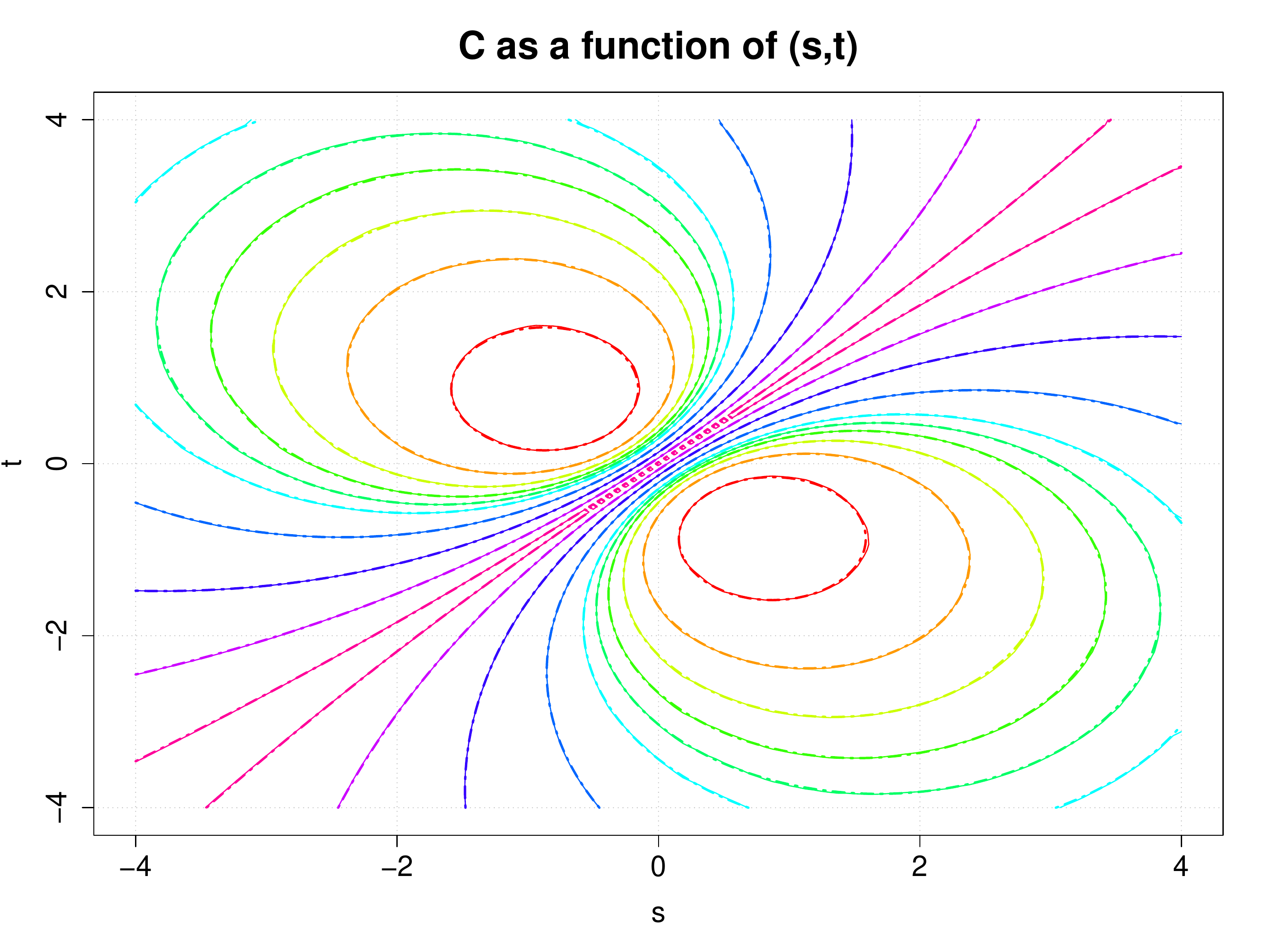}
\caption{A User Defined non stationary covariance function and its estimation from several given fields.}
\label{nonStatCovFunc}
\end{center}
\end{figure}

Any field can be exported into the VTK format which allows it to be visualized using e.g.ParaView (www.paraview.org).\\
Multivariate ARMA stochastic processes $\bm{X}: \Omega \times [0,T] \rightarrow \R^{d}$ are implemented in \ot which enables some manipulations on times series such as the Box Cox transformation or the addition / removal of a trend. Note that the parameters of the Box Cox transformation can be estimated from given fields of the process.\\
\ot models  normal processes, whose  covariance function is a parametric model (e.g. the multivariate Exponential model) as well as  defined by the User as illustrated in Figure \ref{nonStatCovFunc}. Stationary processes can be defined by its spectral density function (e.g. the Cauchy model). 

With explicit notations, the following python commands create a stationary Normal process defined by its covariance function, discretized on a mesh, with an additional trend: 
\lstset{language=Python}
{\small
\begin{lstlisting}
>>>myNormalProcess=TemporalNormalProcess(myTrend, myCovarianceModel, myMesh)
\end{lstlisting}
}

Note that \ot enables the mapping of  any stochastic processes $\bm{X}$ into a process $\bm{Y}$ through a function $f$: $\bm{Y}=f(\bm{X})$ where the function $f$ can consist, for example, of adding or removing a trend, applying a Box Cox transformation in order to stabilize the variance of $\bm{X}$. The python command is, with explicit notations:
{\small
\lstset{language=Python}
\begin{lstlisting}
>>>myYprocess=CompositeProcess(f,myXprocess)
\end{lstlisting}
}

Finally, \ot implements multivariate processes defined as a linear combination of $K$ deterministic functions $(\phi_i)_{i=1,\dots,K}: \R^{d_1} \mapsto \R^{d_2}$:
	\begin{eqnarray*}
	  X(\omega,\bm{x})=\sum_{i=1}^KA_i(\omega)\phi_i(\bm{x})
	\end{eqnarray*}
where $(A_1,\dots, A_K)$ is a random vector of dimension $K$. The python command writes:
{\small
\lstset{language=Python}
\begin{lstlisting}
>>>myX =FunctionalBasisProcess(myRandomCoeff, myBasis, myMesh)
\end{lstlisting}
}

\subsection{Statistics estimation}

\ot enables the User to estimate a model from data, in the univariate as well as in the multivariate framework, using the maximum likelihood principle or the moments based estimation.\\
Some tests, such as the Kolmogorov-Smirnov test, the Chi Square test and the Anderson Darling test (for normal distributions), are implemented and can help to select a model amongst others, from a sample of data. The python command to build a model and test it, writes:
{\small
\lstset{language=Python}
\begin{lstlisting}
>>>estimatedBeta = BetaFactory(sample)
>>>testResult = FittingTest.Kolmogorov(sample, estimatedBeta)
\end{lstlisting}
}

\ot also implements the kernel smoothing technique which is a non-parametric technique to fit a model to data: any distribution can be used as kernel. In the multivariate case, \ot uses the product kernel. It also implements an optimized strategy to select the bandwidth, depending on the number of data in the sample, which is a mix between the Silverman rule and the plugin one. Note that \ot proposes a special treatment when the data are bounded, thanks to the mirroring technique. The python command to build the non-parametric model and to draw its pdf, is as simple as the following one:
{\small
\lstset{language=Python}
\begin{lstlisting}
>>>estimatedDist = KernelSmoothing().build(sample)
>>>pdfGraph = estimatedDist.drawPDF()
\end{lstlisting}
}

\begin{figure}[!ht]
	  \begin{minipage}{7.5cm}
	    \begin{center}
	    \includegraphics[width=7.5cm]{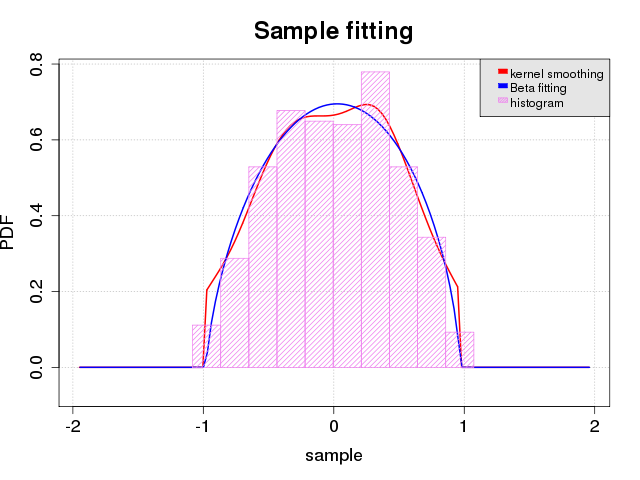}
	    \caption{Beta distribution estimation from a sample of size 500: parametric estimation versus kernel smoothing technique.}
	    \label{betaEstim}
	    \end{center}
	  \end{minipage}
	  \hfill
	  \begin{minipage}{7.5cm}
	    \begin{center}
	       \includegraphics[width=7.5cm]{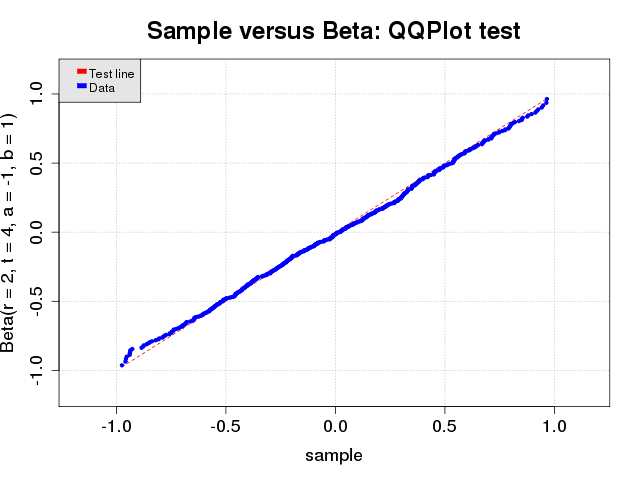}
	       \caption{QQ-plot test: theoretical model Beta versus the sample of size 500.}
	    \label{VisualTest}
	    \end{center}
	  \end{minipage}
	\end{figure}

Figure \ref{betaEstim} illustrates the resulting estimated distributions from a sample of size 500 issued from a Beta distribution: the Kernel Smoothing method takes into account the fact that data are bounded by 0 and 1. The histogram of the data is drawn to enable comparison.

Several visual tests are also implemented to help to select models: among them, the QQ-plot test and  the Henry line test  which writes (in the case of a Beta distributino for the QQ-plot test): 
{\small
\lstset{language=Python}
\begin{lstlisting}
>>>graphQQplot = VisualTest.DrawQQplot(sample, Beta())
>>>graphHenryLine = VisualTest.DrawHenryLine(sample)
\end{lstlisting}
}

Figure \ref{VisualTest} illustrates the QQ-plot test on a sample of size 500  issued from a Beta distribution: the adequation seems satisfying! 

Stochastic processes also have estimation procedures from sample of fields or, if the ergodic hypothesis is verified, from just one field. Multivariate ARMA processes  are estimated according to the BIC and AIC criteria  and the Whittle estimator, which is based on the maximization of the likelihood function in the frequency domain. The python command to estimate an $ARMA(p,q)$ process of dimension $d$, based on a sample of time series, writes:
{\small
\lstset{language=Python}
\begin{lstlisting}
>>>estimatedARMA = ARMALikelihood(p,q,d).build(sampleTimeSeries)
\end{lstlisting}
}

Moreover, \ot can estimate the covariance function and the spectral density function of normal processes from given fields. For example, the python command to estimate a stationary covariance model from a sample of realizations of the process:
{\small
\lstset{language=Python}
\begin{lstlisting}
>>>myCovFunc = StationaryCovarianceModelFactory().build(sampleProcess)
\end{lstlisting}
}
This estimation is illustrated in Figure \ref{nonStatCovFunc}.

\subsection{Conditioned distributions}

\begin{figure}[!ht]
	  \begin{minipage}{7.5cm}
	    \begin{center}
	    \includegraphics[width=7.5cm]{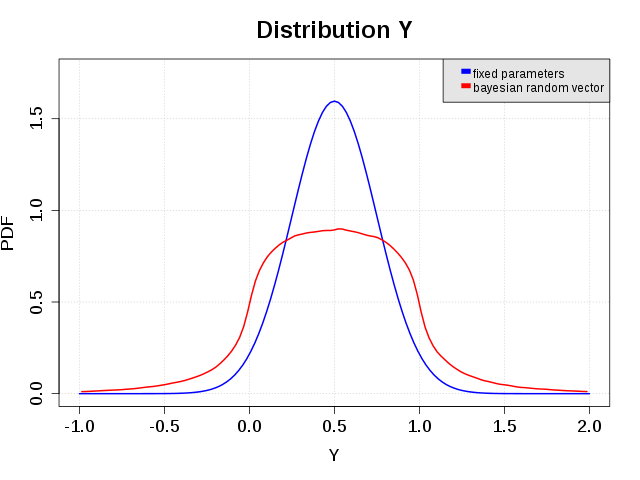}
	    \caption{Normal distribution with random or fixed parameters.}
	    \label{bayesdist}
	    \end{center}
	  \end{minipage}
	  \hfill
	  \begin{minipage}{7.5cm}
	    \begin{center}
	       \includegraphics[width=7.5cm]{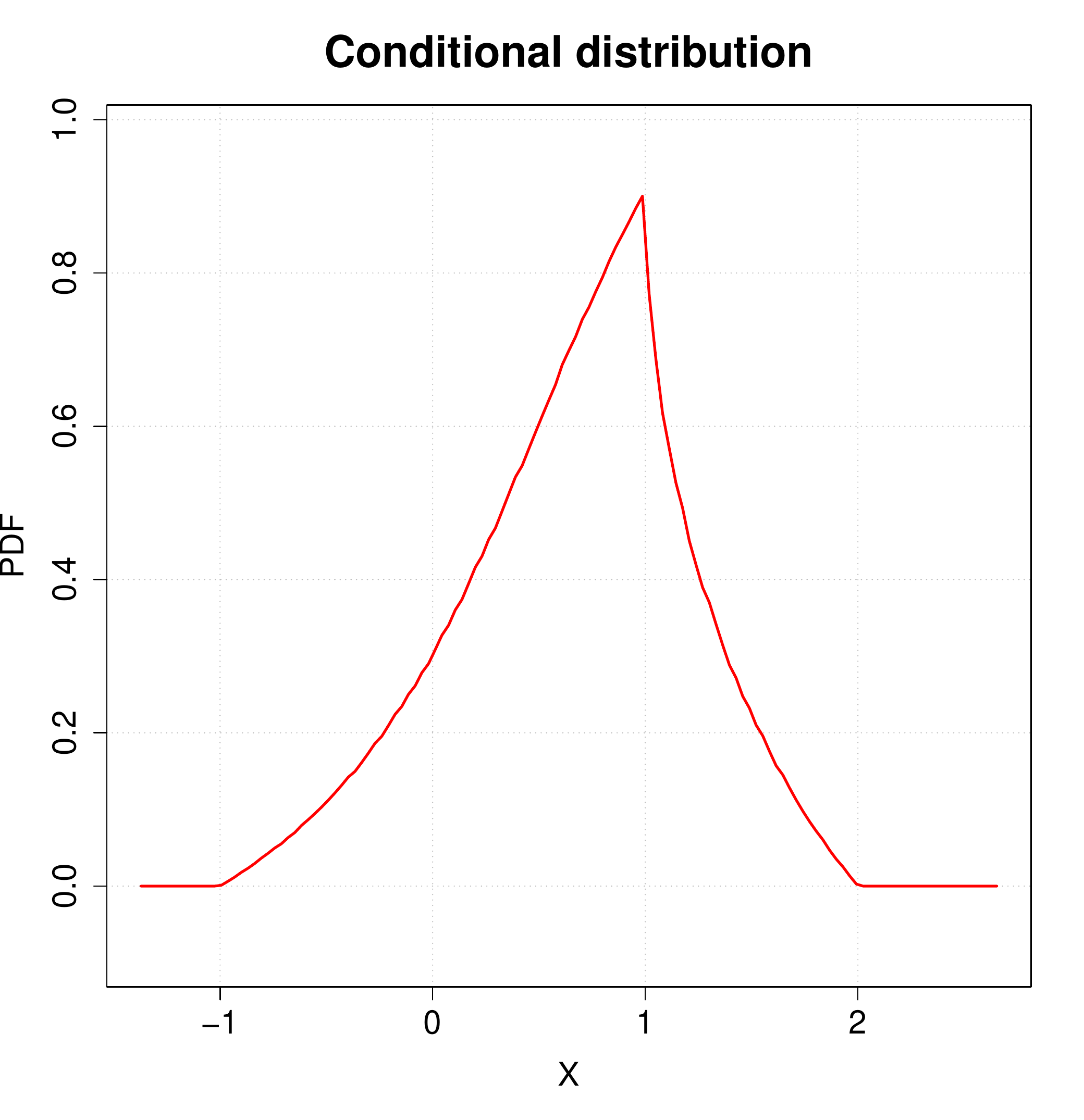}
	       \caption{$Uniform(Y, 1+Y^2)$,  with $Y\sim Uniform(-1,1)$.}
	    \label{ConditionalPDF}
	    \end{center}
	  \end{minipage}
	\end{figure}

\ot enables the modeling of multivariate distributions by conditioning. Several types of conditioning are implemented.\\
At first, \ot enables the creation of a random vector $\bm{X}$ whose distribution $\mathcal{D}_{\bm{X}|\bm{\Theta}}$ whose parameters $\bm{\Theta}$ form a random vector distributed according to the distribution  $\mathcal{D}_{\bm{\Theta}}$. The python command writes:
{\small
\lstset{language=Python}
\begin{lstlisting}
>>>myXrandVect = ConditionalRandomVector(distXgivenTheta, distTheta)
\end{lstlisting}
}
Figure \ref{bayesdist} illustrates a  random variable $X$ distributed according to a Normal distribution: $\mathcal{D}_{\bm{X}|\bm{\Theta}=(M, \Sigma)} = Normal(M, \Sigma)$, which parameters are defined by $M \sim Uniform([0,1])$ and $\Sigma \sim Exponential(\lambda=4)$. The probability density function of $X$  has been fitted with the kernel smoothing technique from $n=10^6$ realizations of $X$ with the normal kernel. It also draws, for comparison needs, the probability density function of $X$ in the case where the parameters are fixed to their mean value.\\
Furthermore, when the random vector  $\bm{\Theta}$ is defined as $\bm{\Theta}=g(\bm{Y})$ where the random vector follows a known distribution $\mathcal{D}_{\bm{Y}}$ and $g$ is a given function, \ot creates the distribution of $\bm{X}$ with the python command:
{\small
\lstset{language=Python}
\begin{lstlisting}
>>>finalDist = ConditionalDistribution(distXGgivenTheta, distY, g)
\end{lstlisting}
}

Figure \ref{ConditionalPDF} illustrates the distribution of  $X$ that follows a $Uniform(A,B)$ distribution, with $(A,B)=g(Y)$, $g:\R \rightarrow \R^2, \, g(Y)=(Y, 1+Y^2)$ and $Y$ follows a $Uniform(-1, 1)$ distribution.\\
	
\subsection{Bayesian Calibration}

Finally, \ot enables  the calibration of a model (which can be a computer code) thanks to the Bayesian estimation, which is the evaluation of the model's parameters. More formally,  let's consider a  model $G$ that writes: $\bm{y} = G(\bm{x}, \bm{\theta})$ where  $\bm{x} \in \R^{d_1}$, $\bm{y} \in \R^{d_3}$ and $\bm{\theta} \in \R^{d_2}$ is the vector of unknown parameters to calibrate. The Bayesian calibration consists in estimating  $\bm{\theta}$, based on a certain set of $n$ inputs $(\bm{x}^1, \ldots, \bm{x}^n)$ (an experimental design) and some associated observations $(\bm{z}^1, \ldots, \bm{z}^n)$ which are regarded  as the realizations of some random vectors $(\bm{Z}^1, \ldots, \bm{Z}^n)$, such that, for all $i$, the distribution of $\bm{Z}^i$ depends on $\bm{y}^i = g(\bm{x}^i, \bm{\theta})$. Typically, $\bm{Z}^i = \bm{Y}^i + \bm{\varepsilon}^i$ where $\bm{\varepsilon}^i$ is a random measurement error. Once the User has defined the prior distribution of $\bm{\theta}$, \ot maximizes the likelihood of the observations and determines the posterior distribution of $\bm{\theta}$, given the observations, using the Metropolis-Hastings algorithm \citep{berger, marin}.

\section{Uncertainty propagation}\label{UncertaintyPropagation}

Once the input multivariate distribution has been satisfactorily chosen, these uncertainties can be propagated through the $G$ model to the output vector $Y$.
Depending on the final goal of the study (min-max approach, central tendency, or reliability), several methods can be used to estimate the corresponding quantity of interest, tending to respect the best compromise between the accuracy of the estimator and the number of calls to the numerical, and potentially costly, model. 

\subsection{Min-Max approach}
The aim here is to determine the extreme (minimum and maximum) values of the components of $Y$ for the set of all possible values of \bm{X}. Several techniques enable it to be done : 
\begin{itemize} 
\item techniques based on design of experiments  : the extreme values of $Y$ are sought for only a finite set of combinations $(\mb{x}^1, \ldots, \mb{x}^n)$, 
\item techniques using optimization algorithms. 
\end{itemize} 

{\bf Techniques based on design of experiments }\\ 
In this case, the min-max approach consists of three steps: 
\begin{itemize} 
\item choice of experiment design used  to determine the combinations  $(\mb{x}^1, \ldots, \mb{x}^n)$ of the input random variables, 
\item evaluation of  $\mb{y}^i = G(\mb{x}^i)$ for $i=1,\ldots,N$, 
\item evaluation of  $\min_{1 \leq i \leq N} y^k_i$  and of  $\max_{1 \leq i \leq N} y^k_i$, together with the combinations related to these extreme values: $\mb{x}_{k,\min} = \textrm{argmin}_{1 \leq i \leq N} y^k_i$  and $\mb{x}_{k,\max} = \textrm{argmax}_{1 \leq i \leq N} y^k_i$. 
\end{itemize} \vspace{2mm} 

The type of design of experiments impacts the quality of the meta model and then on the evaluation of its extreme values. 
\ot gives access to two usual family of design of experiments for a min-max study :
\begin{itemize} 
\item some stratified patterns (axial, composite, factorial or box patterns)
Here are the two command lines that generate a sample from a 2-level factorial pattern. 
{\small
\lstset{language=Python} \begin{lstlisting}
>>>myCenteredReductedGrid = Factorial(2, levels)
>>>mySample = myCenteredReducedGrid.generate()
\end{lstlisting}
}

\item some weighted patterns that include on the one hand,  random patterns (Monte Carlo, LHS), and on the other hand, low discrepancy sequences (Sobol, Faure, Halton, Reverse Halton and Haselgrove, in dimension $n>1$). The following lines illustrates the creation of a Faure sequence in diension 2 or a Monte Carlo design experiments form a bi-dimensional Normal(0,1) distribution. 

{\small
\lstset{language=Python} \begin{lstlisting}
# Sobol Sequence Sampling
>>>mySobolSample = FaureSequence(2).generate(1000)
# Monte Carlo Sampling
>>>myMCSample = MonteCarloExperiment(Normal(2), 100)
\end{lstlisting}
}
\end{itemize} 

\begin{figure}[!ht]
	  \begin{minipage}{7.5cm}
	    \begin{center}
	      \includegraphics[height=5cm]{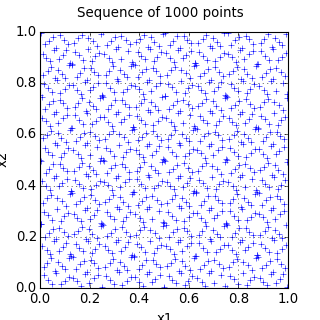}
	      \caption{The first 1000 points according of a Faure sequence of dimension 2.}
	      \label{FaureSeq}
	    \end{center}
	  \end{minipage}
	  \hfill
	  \begin{minipage}{7.5cm}
	    \begin{center}
	      \includegraphics[height=5cm]{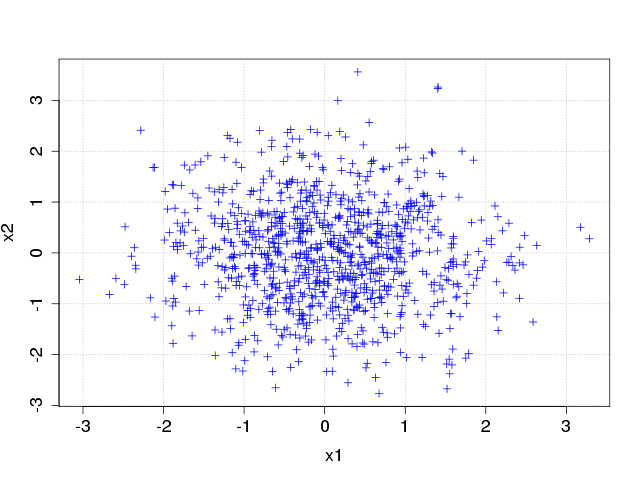}
	      \caption{Sample of 1000 points according of a Normal(0,1) distribution of dimension 2.}
	      \label{MCSeq}
	    \end{center}
	  \end{minipage}
	\end{figure}

Figures \ref{FaureSeq} and \ref{MCSeq} respectively illustrate a design of experiments issued from a Faure sequence or a Normal distribution in dimension 2.

{\bf Techniques based on optimization algorithm}\\ 

In this kind of approach, the min or max value of the output variable is sought thanks to an optimization algorithm. 
\ot offers several optimization algorithms for the several steps of the global methodology. Here the TNC (Truncated Newton Constrainted) is often used, which minimizes a function with variables subject to bounds, using gradient information.
More details may be found in \citep{nash00}.

{\small
\lstset{language=Python} \begin{lstlisting}
# For the research of the min value
>>>myAlgoTNC = TNC(TNCSpecificParameters() , limitStateFunction , 
                   intervalOptim , startingPoint , TNC.MINIMIZATION)
# For the research of the max value
>>>myAlgoTNC = TNC(TNCSpecificParameters() , limitStateFunction , 
                   intervalOptim , startingPoint , TNC.MAXIMIZATION)
# Run the research and extract the results
>>>myAlgoTNC.run()
>>>myAlgoTNCResult = BoundConstrainedAlgorithm(myAlgoTNC).getResult()
>>>optimalValue = myAlgoTNCResult.getOptimalValue()
\end{lstlisting}
}

\subsection{Central tendency}
A central tendency evaluation aims at evaluating a reference value for the variable of interest, here the water level $H$, and an indicator of the dispersion of the variable around the reference. To address this problem, mean $\mu_{Y}=\E(Y)$, and the standard deviation $\sigma_Y=\sqrt{\mathbb{V}(Y)}$ of $Y$ are here evaluated using two different methods. 

First, following the usual method within the Measurement Science community \citep{gum2008}, $\mu_{\mb{Y}}$ and $\sigma_Y$ have been computed under a Taylor first order approximation of the function $Y=G(\mb{X})$ (notice that the explicit dependence on the deterministic variable $\bm{d}$ is here omitted for simplifying notations):
\begin{eqnarray}\label{eq:Taylor1}
	\mu_Y & {\simeq} & G\left(\mathbb{E}(\bm{X})\right) \\
	\sigma_Y & {\approx} & \sum\limits_{i=1}^{d}\sum\limits_{j=1}^{d} \dfrac{\partial G}{\partial \bm{X}_{i}}\Bigr| _{\E(\bm{X})}
	\dfrac{\partial G}{\partial \bm{X_{j}}}\Bigr| _{\E(\bm{X})}\rho_{ij}\sigma _{i}\sigma _{j},
\end{eqnarray}

$\sigma_i$ and $\sigma_j$ being the standard deviation of the ith and jth component ${X}_i$ and ${X}_j$ of the vector \bm{X} and $\rho_{ij}$ their correlation coefficient. Thanks to the formulas above, the mean and the standard deviation of $H$ are evaluated as 52.75m and 1.15m respectively.

{\small
\lstset{language=Python} \begin{lstlisting}
>>>myQuadCum = QuadraticCumul(outputVariable)
# First order Mean
>>>meanFirstOrder =  myQuadCum.getMeanFirstOrder()[0]
# Second order Mean
>>>meanSecondOrder = myQuadCum.getMeanSecondOrder()[0]
# First order Variance
>>>varFirstOrder = myQuadCum.getCovariance()[0,0]
\end{lstlisting}
}

\begin{figure}[ht!]
\centering
\includegraphics[width=0.67\textwidth]{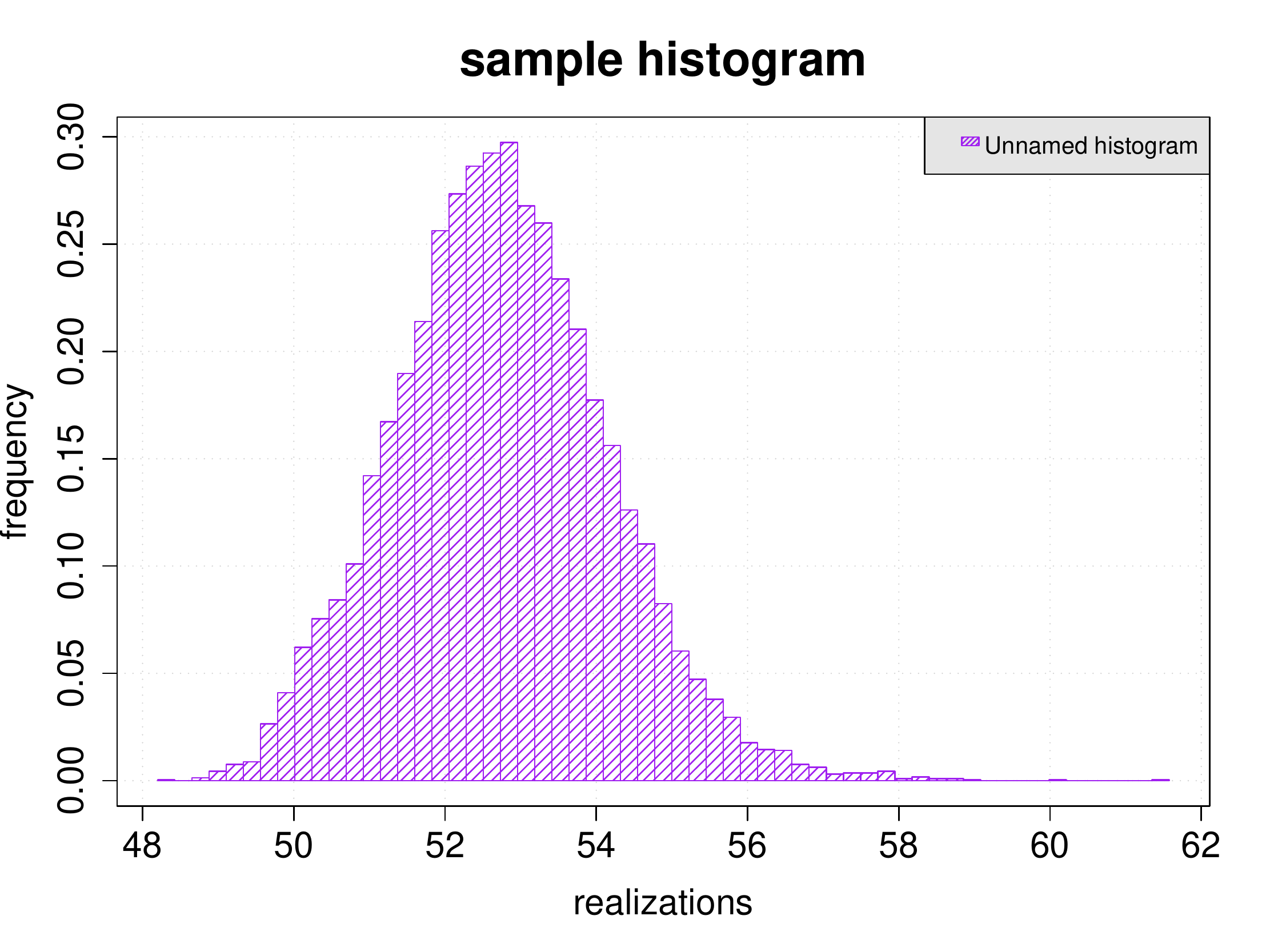}
\caption{Empirical histogram of 10000 samples of $H$.\label{FIGHistogramH}}
\end{figure}

Then, the same quantities have been evaluated by a Monte Carlo evaluation : a set of 10000 samples of the vector $\bm{X}$ is generated and the function $G(\mb{X})$ is evaluated, thus giving a sample of $H$. The empirical mean and standard deviation of this sample are 52.75m and 1.42m respectively.  Figure \ref{FIGHistogramH} shows the empirical histogram of the generated sample of $H$.

{\small
\lstset{language=Python} 
\begin{lstlisting}
# Create a random sample of the output variable of interest of size 10000
>>>outputSample = outputVariable.getNumericalSample(10000)
# Get the empirical mean
>>>empiricalMean = outputSample.computeMean()
# Get the empirical covariance matrix
>>>empiricalCovarianceMatrix = outputSample.computeCovariance()
\end{lstlisting}
}

\subsection{Failure probability estimation}
This section focuses on the estimation of the probability for the output $Y$ to exceed a certain threshold $s$, noted $P_f$ in the following. If $s$ is the altitude of a flood protection dyke, then the above excess probability, $P_f$ can be interpreted as the probability of an overflow of the dyke, i.e. a failure probability. 

Note that an equivalent way of formulating this reliability problem would be to estimate the $(1 - p)$-th quantile of the output's distribution. This quantile can be interpreted as the flood height $q_p$ which is attained with probability $p$ each year. $T = 1/p$ is then seen to be a return period, i.e. a flood as high than $q_{1/T}$ occurs on average every $T$ years. 

Hence, the probability of overflowing a dyke with height $s$ is less than $p$ (where $p$, for instance, could be set according to safety regulations) if and only if $s \geq q_p$, i.e. if the dyke's altitude is higher than the flood with return period equal to $T = 1/p.$

\subsubsection{FORM} \label{sec:excprob}
A  way to evaluate such failure probabilities is through the so-called First Order Reliability Method (FORM) \citep{Ditlevsen1996}. This approach allows, by using an equiprobabilistic transformation and an approximation of the limit-state function, the evaluation with a much reduced number of model evaluations, of some low probability as required in the reliability field. Note that \ot implements the Nataf transformation where the input vector $\bm{X}$ has a normal copula, the generalized Nataf transformation when $\bm{X}$ has an elliptical copula, and the Rosenblatt transformation for any other cases \citep{lebdut09t, lebdut09q, lebdut09b, lebdut09}.

The probability that the yearly maximal water height $H$ exceeds s=58m is evaluated using FORM. The Hasofer-Lind Reliability index was found to be equal to: $\beta_{HL} = 3.04,$ yielding a final estimate of: 
\begin{eqnarray*}
\hat P_{f,FORM} &=& 1.19 \times 10^{-3}.
\end{eqnarray*}

\begin{figure}[ht!]
\centering
\includegraphics[width=0.67\textwidth]{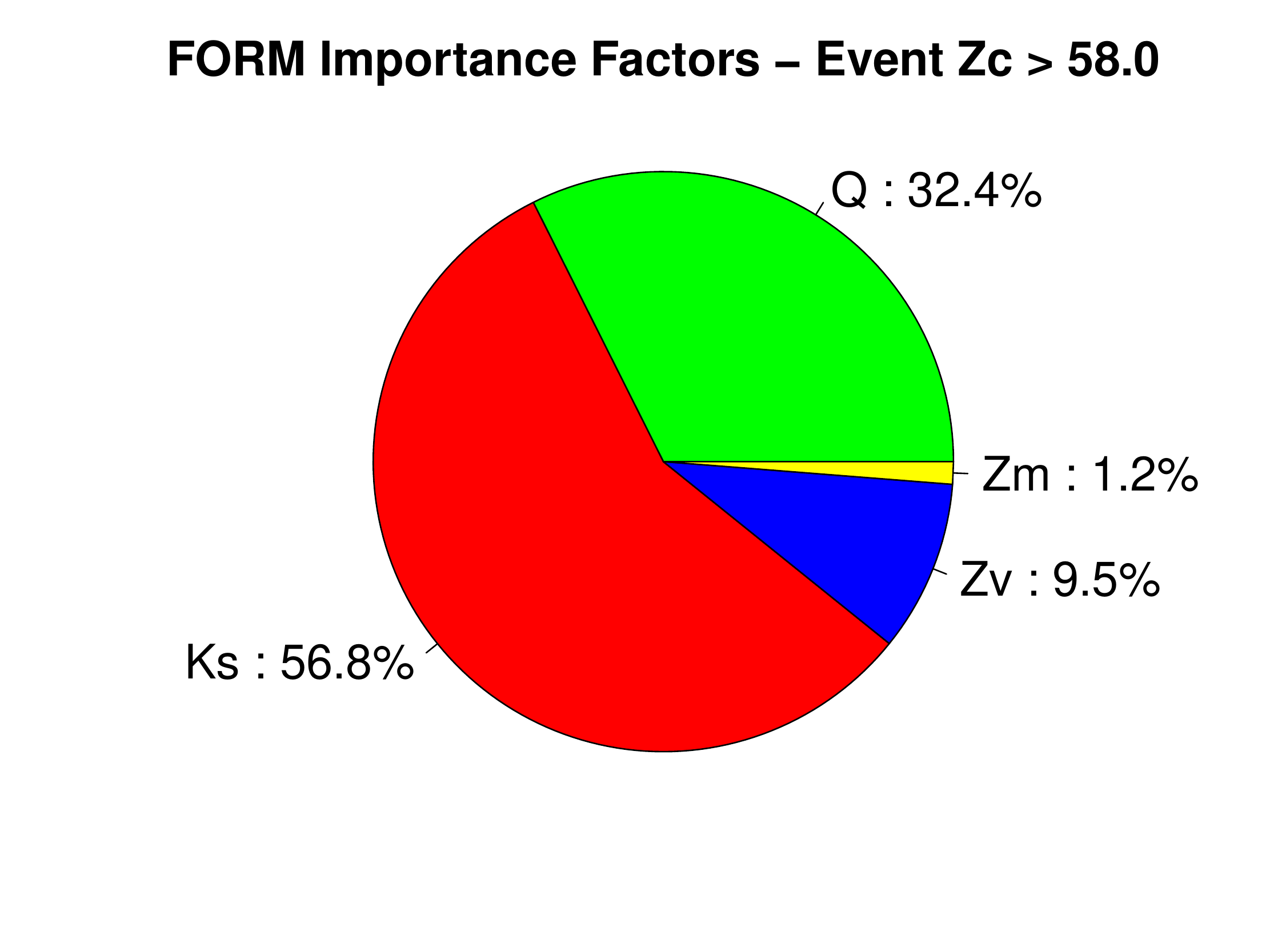}
\caption{FORM Importance Factors.\label{FIGImportanceFactorsFORM}}
\end{figure}

The method gives also some importance factors that measure the weight of each input variable in the probability of exceedance, as shown on Figure \ref{FIGImportanceFactorsFORM}

{\small
\lstset{language=Python} \begin{lstlisting}
>>>myFORM = FORM(Cobyla(), myEvent, meanInputVector)
>>>myFORM.run()
>>>FormResult = myFORM.getResult()
>>>pFORM = FormResult.getEventProbability()
>>>HasoferIndex = FormResult.getHasoferReliabilityIndex()
# Importance factors
>>>importanceFactorsGraph =  FormResult.drawImportanceFactors()
\end{lstlisting}
}

\subsubsection{Monte Carlo}
\begin{figure}[ht!]
\centering
\includegraphics[width=0.67\textwidth]{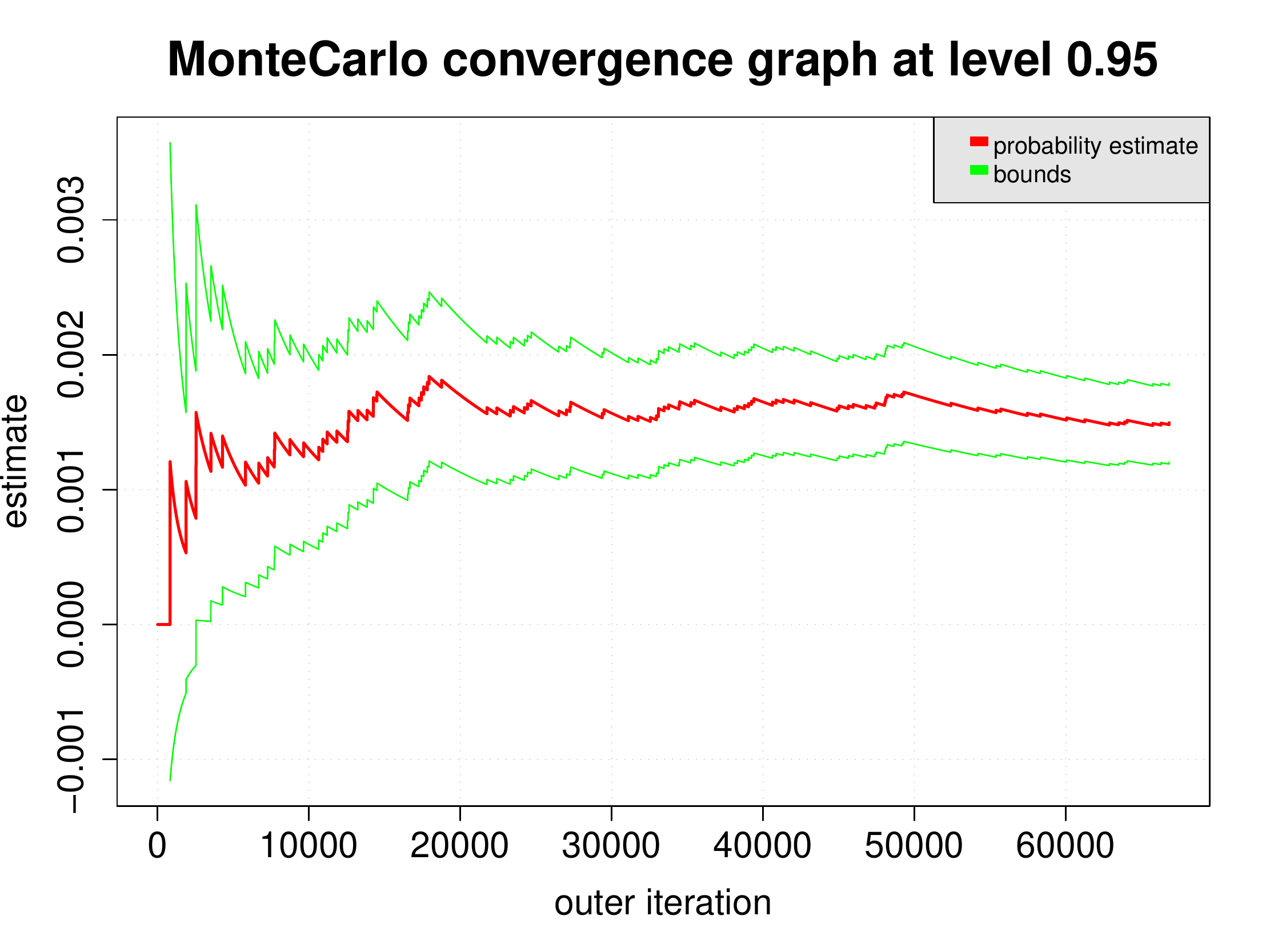}
\caption{Monte Carlo Convergence graph.\label{FIGMonteCarloCovergenceGraph}}
\end{figure}

Whereas the FORM approximation relies on strong assumptions, the Monte Carlo method is always valid, independently from the regularity of the model. It is nevertheless much more computationally intensive, covering all the input domain to evaluate the probability of exceeding a threshold. It consists in sampling many input values $(\bm{X^{(i)}})_{1\leq i \leq N}$ from the input vector joint distribution, then computing the corresponding output values $Y^{(i)} = g(\bm{X^{(i)}}).$ The excess probability $P_f$ is then estimated by the proportion of sampled values $Y^{(i)}$ that exceed $t:$
\begin{eqnarray}
\hat P_{f,MC} &=& \frac{1}{N} \sum_{i=1}^N \boldsymbol 1_{\{Y^{(i)} > s\}}.
\end{eqnarray}
The sample average of the estimation error $\hat P_{f,MC} - P_f$ decreases as $1/\sqrt N,$ and can be precisely quantified by a confidence interval derived from the central limit theorem. In the present case the results are:
\begin{eqnarray*}
\hat P_{f,MC} &=& 1.50 \times 10^{-3},
\end{eqnarray*}
with the following $95\%$ confidence interval:
\begin{eqnarray*}
I_{P_f,MC} &=& \left[ 1.20 \times 10^{-3},  1.79 \times 10^{-3} \right].
\end{eqnarray*}
These results are coherent with those of the FORM approximation, confirming that the assumptions underlying the latter are correct.
Figure \ref{FIGMonteCarloCovergenceGraph} shows the convergence of the estimate depending on the size of the sample, obtained with \ot. 

{\small
\lstset{language=Python} \begin{lstlisting}
>>>myEvent = Event(outputVariable, Greater(), threshold)
>>>myMonteCarlo = MonteCarlo(myEvent)
# Specify the maximum number of simulations
>>>numberMaxSimulation = 100000
>>>myMonteCarlo.setMaximumOuterSampling(numberMaxSimulation)
# Perform the algorithm
>>>myMonteCarlo.run()
# Get the convergence graph
>>>convergenceGraph = myMonteCarlo.drawProbabilityConvergence()
>>>convergenceGraph.draw("MonteCarloCovergenceGraph")
\end{lstlisting}
}

\subsubsection{Importance Sampling}
An even more precise estimate can be obtained through importance sampling \citep{Robert2004}, using the Gaussian distribution with identity covariance matrix and mean equal to the design point $u^\ast$ as the proposal distribution. Many values $(U^{(i)})_{1\leq i\leq N}$ are sampled from this proposal. Because $\phi_n(u - u^\ast)$ is the proposal density from which the $U^{(i)}$ have been sampled, the failure probability can be estimated without bias by:
\begin{eqnarray}
\hat P_{f,IS} &=& \frac{1}{N} \sum_{i=1}^N \boldsymbol 1_{\{G\circ T^{-1}U^{(i)} > s\}} \frac{\phi_n(U^{(i)})}{\phi_n(U^{(i)} - u^\ast)}
\end{eqnarray}
The rationale of this approach is that by sampling in the vicinity of the failure domain boundary, a larger proportion of values fall within the failure domain than by sampling around the origin, leading to a better evaluation of the failure probability, and a reduction in the estimation variance. Using this approach,the results are :
\begin{eqnarray*}
\hat P_{f,IS} &=& 1.40 \times 10^{-3}
\end{eqnarray*}
As in the simple Monte-Carlo approach, a $95\%$-level confidence interval can be derived from the output of the Importance Sampling algorithm. In the present case, this is equal to:
\begin{eqnarray*}
I_{P_f,IS} &=& \left[ 1.26 \times 10^{-3},  1.53 \times 10^{-3} \right],
\end{eqnarray*}
and indeed provides tighter confidence bounds for $P_f$.

{\small
\lstset{language=Python} \begin{lstlisting}
# Specify the starting point from FORM algorithm
>>>standardPoint = FormResult.getStandardSpaceDesignPoint()
# Define the importance distribution 
>>>sigma = [1.0, 1.0, 1.0, 1.0]
>>>importanceDistrib = Normal(standardPoint,sigma,CorrelationMatrix(4))
# Define the IS algorithm : event, distribution, criteria of convergence
>>>myAlgoImportanceSampling = ImportanceSampling (myStandardEvent,importanceDistrib)
>>>myAlgoImportanceSampling.setMaximumOuterSampling(maximumOuterSampling_IS)
>>>myAlgoImportanceSampling.setMaximumCoefficientOfVariation(0.05)
\end{lstlisting}
}

\subsubsection{Directional Sampling}
The directional simulation method is an accelerated sampling method, that involves as a first step a preliminary iso-probabilistic transformation as in FORM method. The basic idea is to explore the space by sampling in several directions in the standard space. The final estimate of the probability $P_f$ after $N$ simulations is the following: 
\begin{align*} 
  \widehat{P}_{f,DS} = \frac{1}{N} \sum_{i=1}^N q_i 
\end{align*} 
where $q_i$ is the probability obtained in each explored direction. A central limit theorem allows to access to some confidence interval on this estimate. More details on this specific method can be found in \citep{Rubinstein81}. 
                                                                               
In practice in \ot, the Directional Sampling simulation requires the choice of several parameters in the methodology : a sampling strategy to choose the explored directions, a "root strategy" corresponding to the way to seek the limit state function (i.e. a sign change) along the explored direction and a non-linear solver to estimate the root. 
A default setting of these parameters allows the user to test the method in one command line : 
{\small
\lstset{language=Python} \begin{lstlisting}
>>>myAlgo = DirectionalSampling(myEvent)
\end{lstlisting}
}

\subsubsection{Subset Sampling}
The subset sampling is a method for estimating rare event probability, based on the idea of replacing rare failure event by a sequence of more frequent events $F_i$.  
\begin{eqnarray*}
F_1 \supset F_2 \supset \dots \supset F_m = F
\end{eqnarray*}
The original probability is obtained conditionaly to the more frequent events : 
\begin{eqnarray*}
P_f = P(F_m) = P(\bigcap \limits_{i=1}^m F_i) = P(F_1) \prod_{i=2}^m P(F_i|F_{i-1})
\end{eqnarray*}
In practice, the subset simulation shows a substantial improvement ($N_T \sim \log P_f$) compared to crude Monte Carlo ($N_T \sim \frac{1}{P_f}$) sampling when estimating rare events.
More details on this specific method can be found in \citep{au_beck01}. 

\ot provides this method through a dedicated module. Here also, some parameters of the methods have to be chosen by the user : a few command lines allows the algorithm to be set up before its launch. 

{\small
\lstset{language=Python} \begin{lstlisting}
>>>mySSAlgo=SubsetSampling(myEvent)
# Change the target conditional probability of each subset domain
>>>mySSAlgo.setTargetProbability(0.9)
# Set the width of the MCMC random walk uniform distribution
>>>mySSAlgo.setProposalRange(1.8)
# This allows to control the number of samples per step
>>>mySSAlgo.setMaximumOuterSampling(10000)
# Run the algorithm
>>>mySSAlgo.run()
\end{lstlisting}
}

\section{Sensitivity analysis}

The sensitivity analysis aims to investigate how a given computational model answers to variations in its inputs. Such knowledge is useful for determining the degree of resemblance of a model and a real system, distinguishing the factors that mostly influence the output variability and those that are insignificant, revealing interactions among input parameters and correlations among output variables, etc. A detailed description of sensitivity analysis methods can be found in \citep{salcha00} and in the \emph{Sensitivity analysis} chapter of the Springer Handbook.
In the global sensitivity analysis strategy, the emphasis is put on apportioning the output uncertainty to the uncertainty in the input factors, given by their uncertainty ranges and probability distributions. 

\subsection{Graphical tools}

\begin{figure}[!ht]
\centering
\includegraphics*[width=13cm]{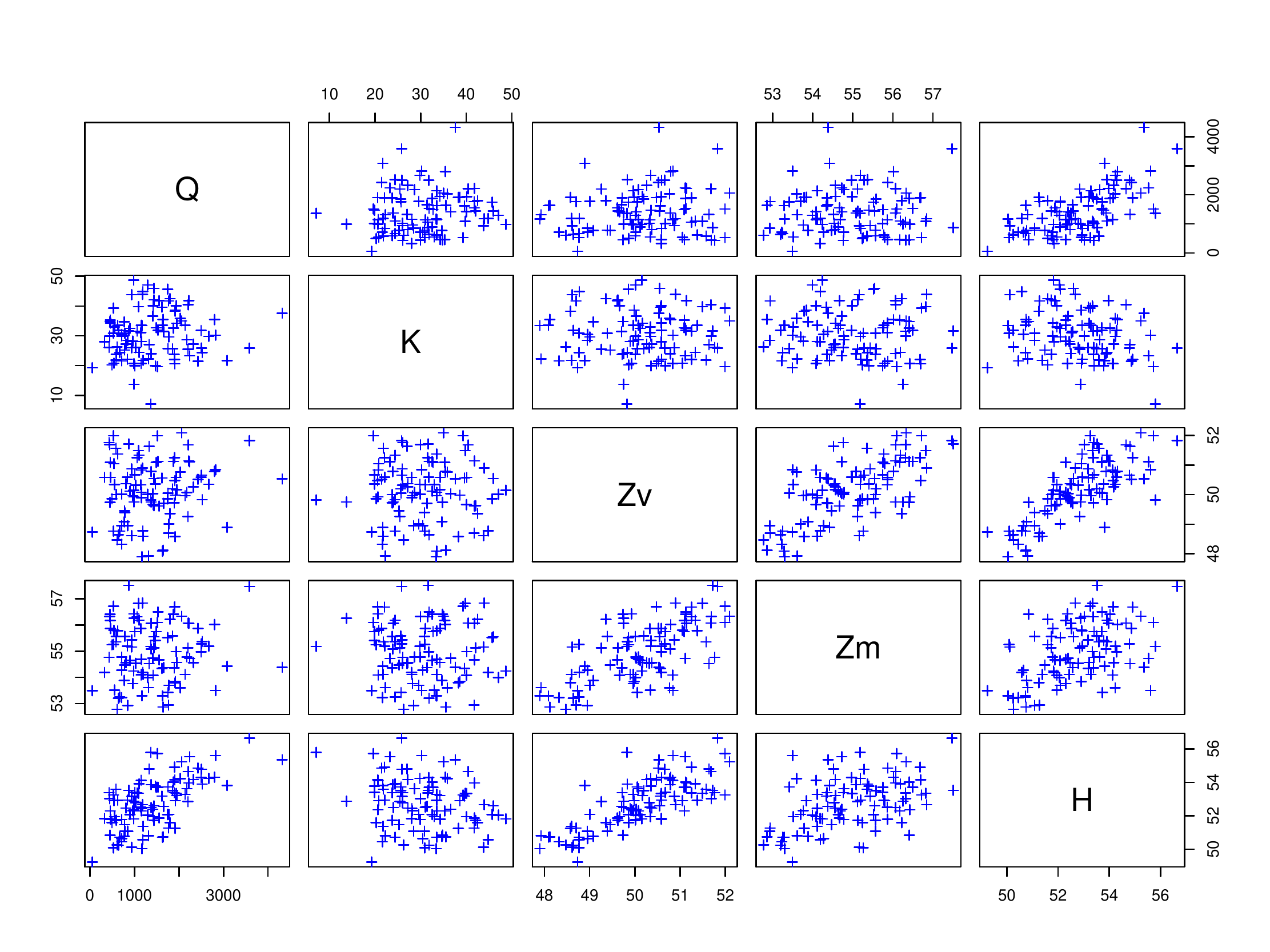}
\caption{Scatterplots between the inputs and the output of the flooding model: each combination (input i, input j) and (input i, output) is drawn, which enables to exhibit some correlation patterns.}
\label{fig:scat}
\end{figure}

In sensitivity analysis, graphical techniques have to be  used first.
With all the scatterplots between each input variable and the model output, one can immediately detect some trends in their functional relation.
The following instructions allow  scatterplots of Figure \ref{fig:scat} to be obtained from a Monte Carlo sample of size $N=1000$ of the flooding model.
{\small 
\lstset{language=Python}
\begin{lstlisting}
>>> inputSample = inputRandomVector.getNumericalSample(1000)
>>> inputSample.setDescription(['Q', 'K', 'Zv', 'Zm'])
>>> outputSample = finalModelCrue(inputSample)
>>> outputSample.setDescription(['H'])
# Here, stack both samples in one
>>> inputSample.stack(outputSample)
>>>myPairs = Pairs(inputSample)
>>>myGraph = Graph()
>>>myGraph.add(myPairs)
\end{lstlisting}}
In the right column of Figure \ref{fig:scat},it is clear that the strong and rather linear effects of $Q$ and $Z_v$ on the output variable $H$.
In the plot of third line and fourth column, it is also clear that the dependence between $Z_v$ and $Z_m$, which comes from the large correlation coefficient introduced in the probabilistic model.

\begin{figure}[!ht]
\centering
\includegraphics*[width=13cm]{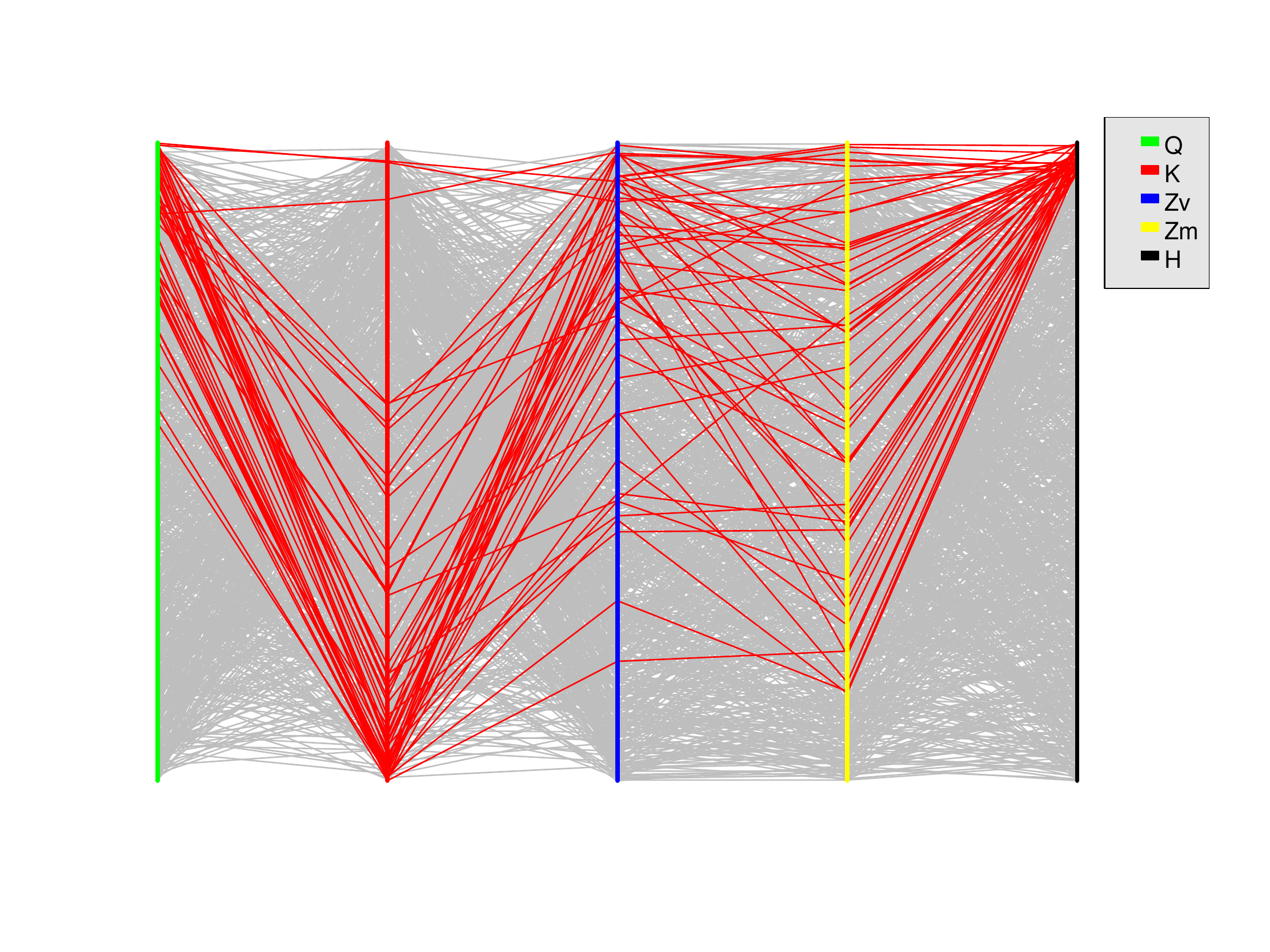}
\caption{Cobweb plot for the flooding model: each simulation is drawn. The input marginal values are linked to the output value (last column). All the simulations that led to a high quantile of the output are drawn in red: the cobweb plot enables to detect all the combinations of the inputs they come from.}\label{fig:cobweb}
\end{figure}

However scatterplots do not capture some interaction effects between the inputs.
Cobweb plots are then used to visualize the simulations as a set of trajectories.
The following instructions allow  the cobweb plots of Figure \ref{fig:cobweb} to be obtained where the simulations leading to the largest values of the model output $H$ have been colored in red.
{\small
\lstset{language=Python}
 \begin{lstlisting}
>>>inputSample = inputRandomVector.getNumericalSample(1000)
>>>outputSample = finalModelCrue(inputSample)
# Graph 1 : value based scale to describe the Y range
>>>minValue = outputSample.computeQuantilePerComponent(0.05)[0]
>>>maxValue = outputSample.computeQuantilePerComponent(0.95)[0]
>>>myCobweb = VisualTest.DrawCobWeb(inputSample, outputSample, 
                                minValue, maxValue, 'red', False)
\end{lstlisting}}
The cobweb plot allows us to immediately understand that these simulations correspond to large values of the flowrate $Q$ and small values of the Strickler coefficient $K_s$.

\subsection{Sampling-based methods}

In order to obtain quantitative sensitivity indices rather than qualitative information, one may use some sampling-based methods which often suppose that the input variables are independent.
The section illustrates some of these methods on the flooding model with independence between its input variables.

If the behavior of the output $Y$ compared to the input vector $\bm{X}$ is overall linear, it is possible to obtain quantitative measurements of the inputs influences from the regression coefficients $\alpha_i$ of the linear regression connecting $Y$ to the $X=(X_1, \ldots, X_{p})$.
The Standard Regression Coefficient (SRC), defined by
\begin{equation}
\mbox{SRC}_i = \alpha_i\frac{\sigma_i}{\sigma_Y} \mbox{ (for $i=1 \ldots p$),}
\end{equation}
with $\sigma_i$ (resp. $\sigma_Y$) the standard deviation of $X_i$ (resp. $Y$),
measures the variation of the response for a given variation of the parameter $X_i$.
In practice, the coefficient $R^2$ (the variance percentage of the output variable $Y$ explained by the regression model) also helps to check the linearity: if $R^2$ is close to one, the relation connecting $Y$ to all the parameters $X_i$ is almost linear and the SRC sensitivity indices make sense.

The following instructions allow  the results of Table \ref{tab:src} to be obtained from a  Monte Carlo sample of size $N=1000$.
{\small 
\lstset{language=Python}
\begin{lstlisting}
>>>inputSample = inputRandomVector.getNumericalSample(1000)
>>>outputSample = finalModelCrue(inputSample)
>>>SRCCoefficient = CorrelationAnalysis.SRC(inputSample, outputSample)
>>>linRegModel=LinearModelFactory().build(inputSample,outputSample,0.90)
>>>Rsquared = LinearModelTest.LinearModelRSquared(inputSample, 
                                         outputSample, linRegModel,0.90)
\end{lstlisting}}
The SRC values confirm our first conclusions drawn from the scatterplots visual analysis.
As $R^2=0.97$ is very close to one, the model is quasi-linear. The SRC coefficients are sufficient to perform a global sensitivity analysis.

\begin{table}[ht!]
\caption{Regression coefficients and SRC of the flood model inputs ($\alpha_0=-0.1675$ and $R^2=0.97$).}\label{tab:src}
\centering
\begin{tabular}{c|c|c|c|c|c}
& $Q$ & $K_s$ & $Z_v$ & $Z_m$ &  \\
\hline
$\alpha_i$ & 3.2640 & 0.0012 & -0.0556 & 1.1720  \\
\hline
$\mbox{SRC}_i$ & 0.3462 & 0.0851 & 0.6814 & 0.0149 
\end{tabular}
\end{table}

Several other estimation methods are available in OpenTURNS for a sensitivity analysis purpose: 
\begin{itemize}
\item derivatives and Pearson correlation coefficients (linearity hypothesis between output and inputs), 
\item Spearman correlation coefficients and Standard Rank Regression Coefficients (monotonicity hypothesis between output and inputs),
\item reliability importance factors with the FORM/SORM importance measures presented previously (section \label{UncertaintyPropagation}),
\item variance-based sensitivity indices (no hypothesis on the model).
These last indices, often known as Sobol' indices and defined by
\begin{equation}
S_i = \frac{\mbox{Var}[\mathbb{E}(Y|X_i)]}{\mbox{Var}(Y)} \mbox{ (first order index) and } S_{T_i} = \sum_{i=1}^p S_i +  \sum_{i<j} S_{ij} + \ldots \mbox{ (total index),}
\end{equation}
are estimated in OpenTURNS with the classic pick-freeze method based on two independent Monte Carlo samples \citep{sal02}.
In OpenTURNS, other ways to compute the Sobol' indices are the Extended FAST method \citep{saltar99} and the coefficients of the polynomial chaos expansion \citep{sud08}.
\end{itemize}

\section{Metamodels}


 When each model evaluation is time consuming, it is usual to build a surrogate model which is a good approximation of the initial model and which can be evaluated  at negligible cost. 
\ot proposes some usual polynomial approximations: the linear or quadratic Taylor approximations of the model at a particular point, or a linear approximation based on the least squares method. 
Two recent techniques implemented in \ot are detailed here: the  polynomial chaos expansion and the kriging approximation.

\subsection{Polynomial chaos expansion}

	The polynomial chaos expansion enables the approximation of the output random variable of interest $\bm{Y} = G(\bm{X})$ with $g: \R^d \rightarrow \R^p$ by the  surrogate model:
	\begin{align*}
	\tilde{\bm{Y}} = \sum_{k \in K} \bm{\alpha}_k \Psi_k \circ T(\bm{X})
	\end{align*}
	where $ \bm{\alpha}_k \in \R^p$, $T$ is an isoprobabilistic transformation (e.g. the Rosenblatt transformation) which maps the multivariate distribution of $\bm{X}$ into the multivariate distribution $\mu = \prod_{i=1}^d \mu_i$, and $(\Psi_k)_{k \in \N}$ a multivariate polynomial basis of $\mathcal{L}^2_{\mu}(\R^d,\R^p)$ which is orthonormal according to the distribution $\mu$. $K$ is a finite subset of $\N$. $\bm{Y}$ is supposed to be of finite second moment.\\
\ot proposes the building of the  multivariate orthonornal basis $(\Psi_k(\bm{x}))_{k \in \N}$ as the cartesian product of orthonormal univariate polynomial family $(\Psi_{l}^i(z_i))_{l \in \N}$ :
	\begin{align*}
	  \Psi_k(\bm{z}) = \Psi_{k_1}^1(z_1) * \Psi_{k_2}^2(z_2) * \dots *  \Psi_{k_d}^d(z_d)
	\end{align*}
	
The possible univariate polynomial families associated to continuous measures are :
	\begin{itemize}
	\item Hermite, which is orthonormal with respect to the  $Normal(0,1)$ distribution,
	\item Jacobi($\alpha$, $\beta$, \textit{param}), which is orthonormal with respect to the $Beta(\beta + 1, \alpha + \beta + 2, -1, 1)$ distribution if $param = 0$ (default value) or to the $Beta(\alpha, \beta, -1, 1)$ distribution if $param = 1$,
	\item Laguerre($k$), which is orthonormal with respect to the  $Gamma(k+1,1,0)$ distribution,
	\item Legendre, which is orthonormal with respect to the  $Uniform(-1,1)$ distribution.
	\end{itemize}

\ot proposes three strategies to truncate the multivariate orthonormal basis to the finite set $K$ : these strategies select different terms from the multivariate basis, based on a convergence criterion of the surrogate model and  the cleaning of the less significant coefficients.\\
The coefficients of the polynomial decomposition writes: 
\begin{equation}\label{defArgMin}
	  \bm{\alpha} = argmin_{\bm{\alpha} \in \R^K} E_{\mu} \left[ \left( g \circ T^{-1}(\bm{Z}) -  \sum_{k \in K} \alpha_k \Psi_k (\bm{Z})\right)^2  \right]
	\end{equation}
	as well as:
	\begin{equation}\label{defEsp}
	  \bm{\alpha} = \left( E_{\mu} \left[ g \circ T^{-1}(\bm{Z}) \Psi_k (\bm{Z}) \right]\right)_k
	\end{equation}
	where $\bm{Z} = T(\bm{X})$ is distributed according to $\mu$.\\
It corresponds to two points of view implemented by \ot: the relation (\ref{defArgMin})  means that the coefficients $(\alpha_k)_{k \in K}$ minimize the mean quadratic error between  the model and the polynomial approximation;  the relation (\ref{defEsp}) means that  $\alpha_k$ is the scalar product of the model  with the $k-th$ element of the orthonormal basis $(\Psi_k)_{k \in K}$. In both cases, the expectation $ E_{\mu}$ is approximated by a linear quadrature formula that writes, in the general case:
	\begin{equation}\label{approxEsp}
	  E_{\mu} \left[ f(\bm{Z}) \right] \simeq \sum_{i \in I} \omega_i f(\Xi_i)
	\end{equation}
where $f$ is a function $L_1(\mu)$. The set $I$, the points $(\Xi_i)_{i \in I}$ and the weights $(\omega_i)_{i \in I}$ are evaluated from weighted designs of experiments which can be random (Monte Carlo experiments, and Importance sampling experiments) or deterministic (low discrepancy experiments, User given experiments, and Gauss product experiments).\\
At last, \ot gives access to:
\begin{itemize}
	\item the composed model $h : \bm{Z} \mapsto \bm{Y} = G \circ T^{-1}(\bm{Z})$, which is the model of the reduced variables $\bm{Z}$. Then  $\displaystyle h =  \sum_{k \in \N} \bm{\alpha}_k \Psi_k$,
	\item the coefficients of the polynomial approximation : $(\bm{\alpha}_k)_{k \in K}$,
	\item the composed meta model: $\hat{h}$, which is the model of the reduced variables reduced to the truncated multivariate basis $(\Psi_k)_{k \in K}$. Then $\displaystyle  \hat{h} = \sum_{k \in K} \bm{\alpha}_k \Psi_k$,
	\item the  meta model: $\displaystyle \hat{g} : \bm{X} \mapsto Y = \hat{h} \circ T(\bm{X})$ which is the polynomial chaos approximation as a NumericalMathFunction. Then $\displaystyle \hat{g} = \sum_{k \in K} \bm{\alpha}_k \Psi_k \circ T$.
\end{itemize}

When the model is very expensive to evaluate, it is necessary to optimize the number of coefficients of the polynomials chaos expansion to be calculated. Some specific strategies have been proposed by \citep{Blatman} for enumerating the infinite polynomial chaos series: \ot implements the hyperbolic enumeration strategy which is inspired by the so-called sparsity-of-effects principle. This strategy states that most models are principally governed by main effects and low-order interactions. This enumeration strategy  selects the multi-indices related to main effects.

The following lines illustrates the case where the model $\mathcal{G}: \bm{x} \mapsto x\sin x$ and where the input random vector follows a Uniform distribution on $[-1, 1]$
{\small
\lstset{language=Python}
\begin{lstlisting}
# Define the model
>>> model = NumericalMathFunction(['x'], ['x*sin(y)'])
# Create the input distribution
>>> distribution = Uniform()
# Construction of the  orthonormal basis
>>> polyColl = [0.]
>>> polyColl[0] = StandardDistributionPolynomialFactory(distribution)
>>> enumerateFunction = LinearEnumerateFunction(1)
>>> productBasis = OrthogonalProductPolynomialFactory(polyColl, enumerateFunction)
# Truncature strategy of the multivariate orthonormal basis
# Choose all the polynomials of degree <= 4
>>> degree = 4
>>> indexMax = enumerateFunction.getStrataCumulatedCardinal(degree)
# Keep all the polynomials of degree <= 4
# which corresponds to the 5 first ones
>>> adaptiveStrategy = FixedStrategy(productBasis, indexMax)
# Evaluation strategy of the approximation coefficients
>>> samplingSize = 50
>>> experiment = MonteCarloExperiment(samplingSize)
>>> projectionStrategy = LeastSquaresStrategy(experiment)
# Creation of the Functional Chaos Algorithm
>>> algo = FunctionalChaosAlgorithm(model, distribution, adaptiveStrategy,
...                                    projectionStrategy)
>>> algo.run()
# Get the result
>>> functionalChaosResult = algo.getResult()
>>> metamodel = functionalChaosResult.getMetaModel()
\end{lstlisting}
}

\begin{figure}[!ht]
\begin{center}
\includegraphics[width=7cm]{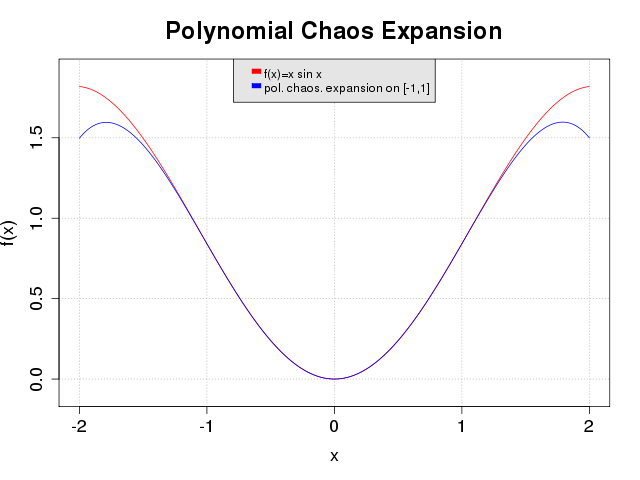}
\caption{An example ofa polynomial chaos expansion: the blue line is the reference function $\mathcal{G}: \bm{x} \mapsto x\sin x$ and the red one its approximation only valid on $[-1, 1]$.}
\label{chaos}
\end{center}
\end{figure}

Figure \ref{chaos} illustrates the result.

\subsection{The Kriging approximation}

 Kriging (also known as Gaussian process regression) \citep{sacwel89, Santner, Rasmussen,marioo08} is a Bayesian technique
that aims at approximating functions (most often in order to surrogate them
because they are expensive to evaluate). In the following it is assumed the aim is to surrogate a scalar-valued model $G: \bm{x} \mapsto y$. Note the
\ot implementation of Kriging can deal with vector-valued functions
($G: \bm{x} \mapsto \bm{y}$), with simple loops over each output.
It is also assumed the model was run over a design of experiments in order
to produce a set of observations gathered in the following dataset:
$\left(\left(\bm{x}^{i}, y^{i}\right), i = 1, \ldots, n\right)$.
Ultimately Kriging aims at producing a predictor (also known as a response
surface or metamodel) denoted as $\tilde{G}$.\\
It is assumed that the model $G$ is a realization of the normal process $Y: \Omega \times \R^d \rightarrow \R$ defined by:
\begin{equation}
    Y(\omega, \bm{x}) = m(\bm{x}) + Z(\omega, \bm{x})
\end{equation}
where $m(\bm{x})$ is the trend and  $Z(\bm{x})$ is a zero-mean Gaussian process with a  covariance function $c_{\bm{\theta}}: \R^d \times \R^d \rightarrow \R$ which depends on the vector of parameters $\bm{\theta} \in \R^{n_\theta}$:
\begin{equation}
    \EE{[Z(\bm{x}), Z(\bm{y})]} = c_{\bm{\theta}}(\bm{x},\bm{y})
\end{equation}

\begin{figure}[!ht]
\begin{center}
\includegraphics[width=7cm]{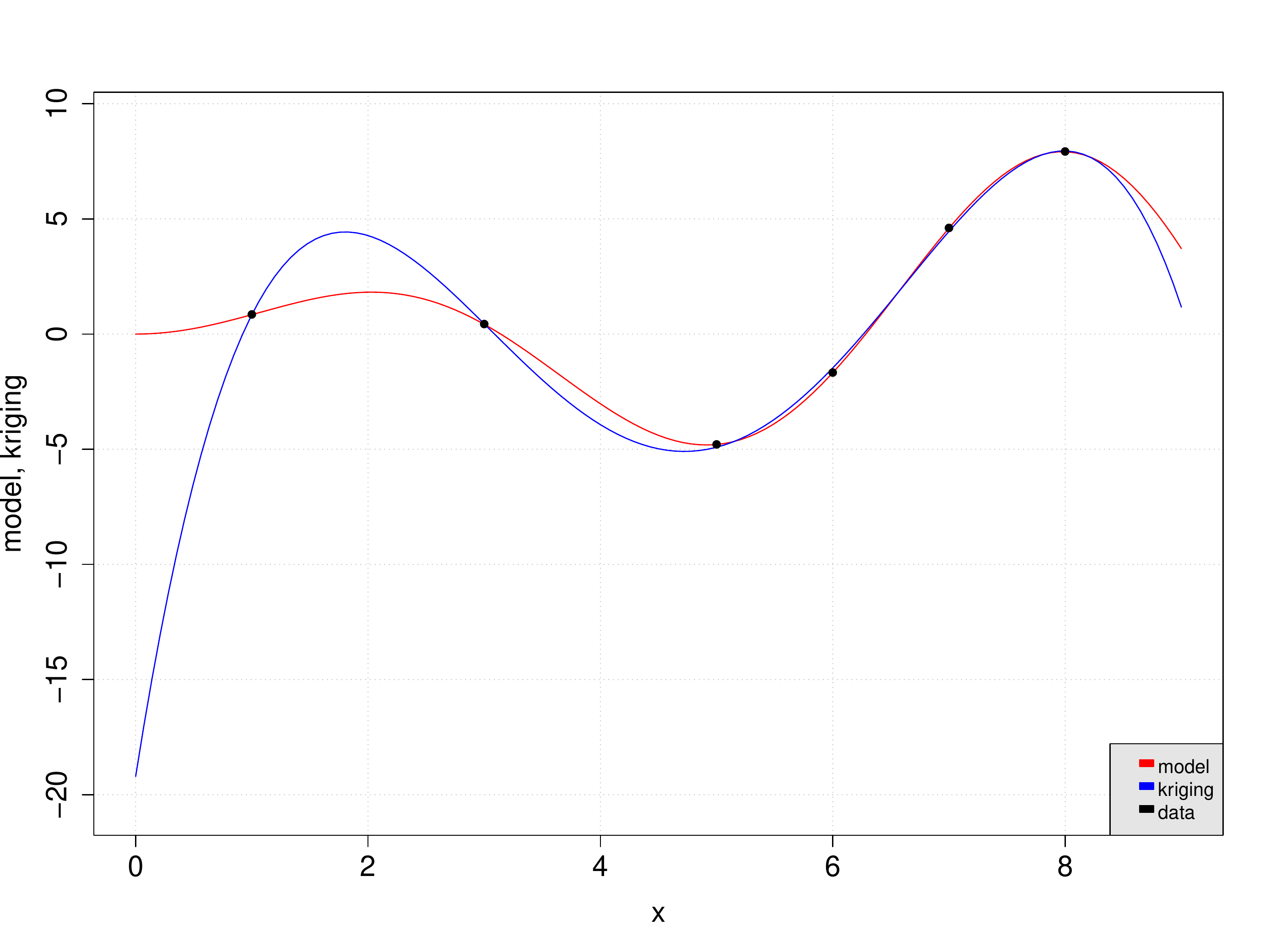}
\caption{An example of kriging approximation based on 6 observations: the blue line is the reference function $\mathcal{G}: \bm{x} \mapsto x\sin x$ and the red one its approximation by a realization of a Gaussian process.}
\label{krigEx}
\end{center}
\end{figure}

The trend is generally taken equal to the generalized linear model:
\begin{equation}
    m(\bm{x}) = \left(\bm{f}(\bm{x})\right)^t \bm{\beta}
\end{equation}  
where $\left(\bm{f}(\bm{x})\right)^t = \left(f_1, \ldots, f_p\right)$
and $\bm{\beta} = \left(\beta_1, \ldots, \beta_p\right)$. Then, the Kriging method approximates the model $f$ by the mean of the $Y$ given that:
\begin{equation}
    Y(\omega, \bm{x}^{(i)}) = y^{(i)} \quad \forall i=1, \dots, n
\end{equation}
The Kriging meta model $\tilde{G}$ of $G$ writes:
\begin{equation}
   \tilde{G}(\bm{x}) = \EE{[Y(\omega, \bm{x})|Y(\omega, \bm{x}^{(i)}) = y^{(i)}, \forall i=1, \dots, n]}
\end{equation}   
The meta model is then defined by:
\begin{equation}
   \tilde{G}(\bm{x}) = \left(\bm{f}(\bm{x})\right)^t\tilde{\bm{\beta}} + \left(\bm{c}_{\bm{\theta}}(\bm{x})\right)^tC_{\bm{\theta}}^{-1}(\bm{y}-F\tilde{\bm{\beta}})
\end{equation}
where $\tilde{\bm{\beta}}$ is the least squares estimator for $\bm{\beta}$ defined by:
\begin{equation}
 \tilde{\bm{\beta}} =  \left(F^tC_{\bm{\theta}}^{-1}F\right)^{-1}F^tC_{\bm{\theta}}^{-1}\bm{y}
\end{equation}
and $C_{\bm{\theta}}=[c_{\bm{\theta}}(x_i,x_j)]_{i,j=1\dots n}$, $F=[f(x_i)^t]_{i=1\dots n}$ and $\bm{c}^t_{\bm{\theta}}(\bm{x})=[c_{\bm{\theta}}(x,x_i)]_{i=1\dots n}$.
The line command writes:
{\small
\lstset{language=Python}
\begin{lstlisting}
>>> algo = KrigingAlgorithm(inputSample, outputSample, basis, covarianceModel)
>>> algo.run()
>>> result = algo.getResult()
>>> metamodel = result.getMetaModel()
>>> graph = metamodel.draw()
\end{lstlisting}
}
Figure \ref{krigEx} approximates the  previously defined model $\mathcal{G}: \bm{x} \mapsto x\sin x$ with a realization of a Gaussian process  based on 6 observations.

\section{Using an external code}

\section{The external simulator}

\subsection{Fast evaluations of G}

On the practical side, the \ot{} software provides features 
which make the connection to the simulator G easy, 
and make its evaluation generally fast.
Within the \ot{} framework, the method to connect to G is 
called "wrapping". 

In the simplest situations, the function G is analytical 
and the formulas can be provided to \ot{} with a character string. 
Here, the Muparser C++ library \cite{MuParserLib} is used to evaluate 
the value of the mathematical function. 
In this case, the evaluation of G by \ot{} is quite fast. 

In the following Python script,  consider the function $G:\R^3\rightarrow \R^2$, 
where $G_1(\mb{x})=x_1+x_2+x_3$ and $G_2(\mb{x})=x_1-x_2x_3$, 
for any real numbers $x_1$, $x_2$ and $x_3$. 
The input argument of the \pyobj{NumericalMathFunction} class is a 
Python \pyobj{tuple}, where the first item describes the 
three inputs variables, the second item describes the two output 
variables and the last item describes the two functions $G_1$ and 
$G_2$. 
\lstset{language=Python}
\small{
\begin{lstlisting}
>>>G = NumericalMathFunction(
    ("x0","x1","x2"),
    ("y0","y1"),
    ("x0+x1+x2","x0-x1*x2"))
\end{lstlisting}
}
Once created, the function \pyobj{G} can be used as a regular 
Python function, or can be passed as an input argument of other 
\ot{} classes. 

In most cases, the function G is provided as a Python function, which 
can be connected to \ot{} with the \pyobj{PythonFunction} class. 
This task is easy (for those who are familiar with this language), 
and allows the scientific packages already available in Python to be combined. 
For example, if the computational code uses XML files on input or output, 
it is easy to make use of the XML features of Python (e.g. the \pyobj{minidom} 
package). 
Moreover, if the function evaluation can be vectorized (e.g. with the 
\pyobj{numpy} package), then the \pyobj{func\_sample} option of the 
\pyobj{PythonFunction} class can improve the performance a lot.

The following Python script creates the function $G$ 
associated with the flooding model. 
The \pyobj{flood} function is first defined with the 
\pyobj{def} Python statement. 
This function takes the variable \pyobj{X} as input argument, 
which is an array with four components, \pyobj{Q}, \pyobj{K\_s}, 
\pyobj{Z\_v} and \pyobj{Z\_m}, which corresponds to the 
input random variables in the model. 
The body of the \pyobj{flood} function is a regular Python 
script, so that all Python functions can be used at this 
point (e.g. the \pyobj{numpy} or \pyobj{scipy} functions). 
The last statement of the function returns the overflow \pyobj{S}. 
Then the \pyobj{PythonFunction} class is used in order to convert this 
Python function into an object that \ot{} can use. 
This class takes as input arguments the number of input 
variables (in this case, 4), the 
number of outputs (in this case, 1) and the 
function and returns the object \pyobj{G}.
\lstset{language=Python}
\small{
\begin{lstlisting}
>>>from openturns import PythonFunction
>>>def flood(X) :
    L = 5.0e3;     B = 300.0
    Q, K_s, Z_v, Z_m = X
    alpha = (Z_m - Z_v)/L
    H = (Q/(K_s*B*sqrt(alpha)))**(0.6)
    return H
>>>G = PythonFunction(4, 1, flood) 
\end{lstlisting}
}

If, as many of the computational codes commonly used, 
the data exchange is based on text files, \ot provides a component 
(\pyobj{coupling\_tools}) which is able to read and write structured text files 
based, for example, on line indices and perhaps containing tables (using 
line and column indices). 
Moreover, \ot provides a component which can evaluate such a Python 
function using the multi-thread capabilities that most computers have.  

Finally, when the computational code G is provided as a C or Fortran library, 
\ot provides a generic component to exchange data by memory, which is much faster 
than with files. 
In this case, the evaluation of G is automatically multi-thread. 
This component can be configured by Python, based on a XML file. 
If this method is not flexible enough, then the connection can be done 
with the C++ library. 

The previous techniques are documented in the \ot Developer's Guide \cite{OTDevGuide14}. 

\begin{figure}
\centering
\includegraphics[width=0.8\textwidth]{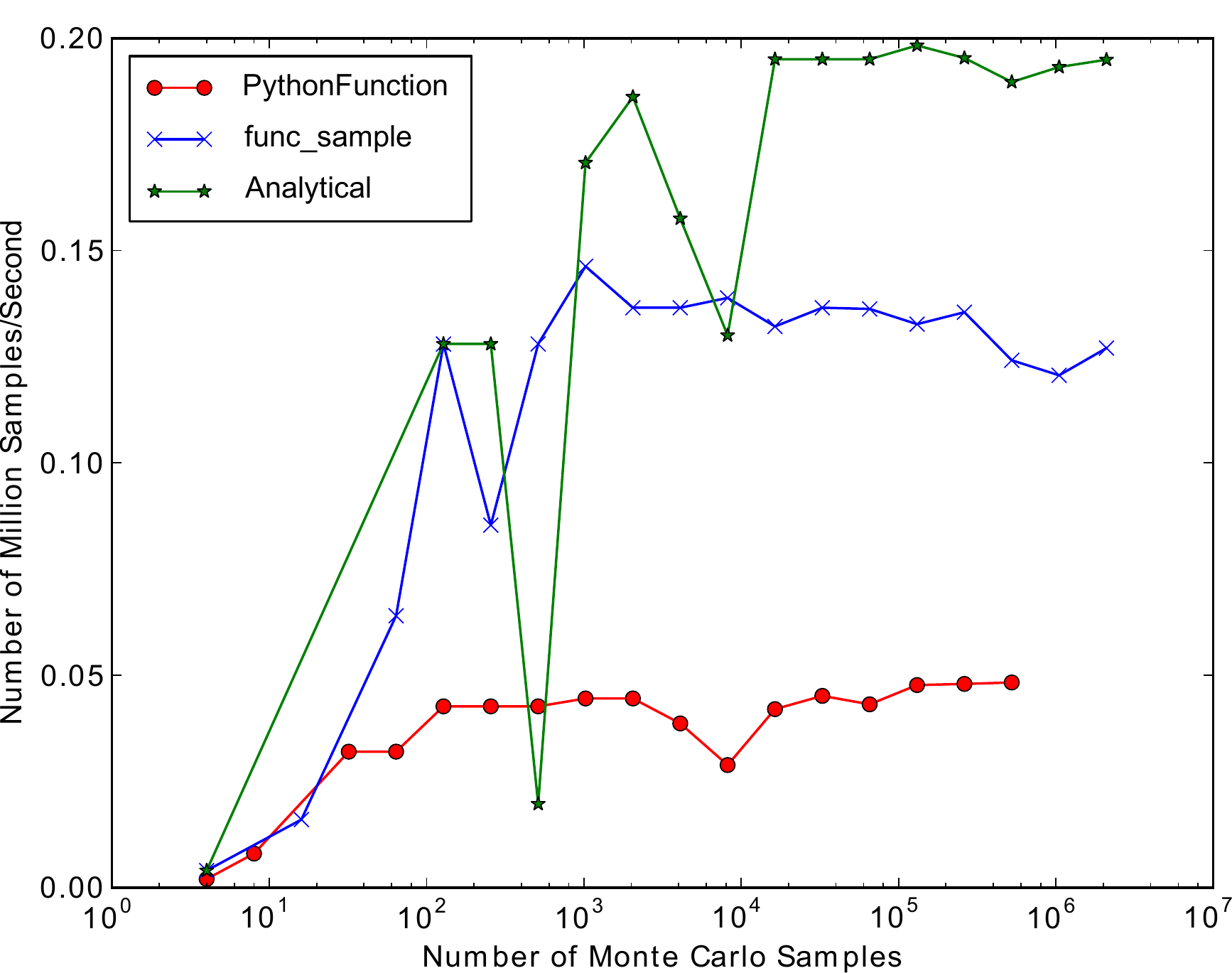}
\caption{Performance of various connection methods in \ot}
\label{fig-benchmarkperf}
\end{figure}

Figure \ref{fig-benchmarkperf} compares the performance of three methods to 
connect to the function G : the \pyobj{PythonFunction} class, the 
\pyobj{PythonFunction} class with the \pyobj{func\_sample} option and the 
analytical function. 
This test was performed with a basic MS Windows laptop computer. 
Obviously, the fastest method is the analytical function, which can provide 
as many as 0.2 million samples per second, a performance which is four times 
the performance of the \pyobj{PythonFunction} class. 

%
%

\subsection{Evaluation of the derivatives}

When the algorithm involves an optimization step (e.g. in the FORM-SORM 
method) or a local approximation of G (e.g. in the Taylor development used 
to approximate the expectation and variance), the derivatives 
of G are required. 

When the computer code can compute the gradient and Hessian matrix 
of G, this information can be used by \ot. 
This happens sometimes, for example when the computer code 
has been differentiated with automatic differentiation methods, 
such as forward or adjoint techniques. 

In the case where the function is analytical and 
is provided as a character string, 
\ot is able to compute the exact derivatives of G. 
In order to do this, the software uses the Ev3 C++ library \cite{Ev3Lib}
to perform the symbolic computation of the derivatives 
and MuParser \cite{MuParserLib} to evaluate it.

In most common situations, however, the code does not compute 
its derivatives. 
In this case, \ot provides a method to compute the 
derivatives based on finite difference formulas. 
By default a centered finite difference formula for the 
gradient and a centered formula for the Hessian matrix are used. 

%
%

\subsection{High performance computing}

For most common engineering practices, \ot can evaluate G with the multi-thread 
capabilities of most laptop and scientific workstations. 
However, when the evaluation of G is more CPU consuming or 
when the number of evaluations required is larger, these
features are not sufficient by themselves and it is necessary to use 
a high performance computer such as the Zumbrota, Athos or 
Ivanhoe supercomputers available at EDF R\&D which have from 16 000 
to 65 000 cores \cite{ZumbrotaTop500}. 

In this case, two solutions are commonly used. 
The first one is to use a feature which can execute a Python function 
on remote processors, connected on the network with \pyobj{ssh}. 
Here, the data flow is based on files, located in automatically generated 
directories, which prevents the loss of intermediate data. 
This feature (the \pyobj{DistributedPythonFunction}) allows each 
remote processor to use its multi-thread capabilities, providing 
two different levels of parallel computing. 

The second solution is to use the \ot component integrated in the 
Salome platform. 
This component, along with a graphical user interface, called "Eficas", 
makes use of a software, called "YACS", which can call a Python script. 
The YACS module allows  calculation schemes in Salome to be built, edited and executed.
It provides both a graphical user interface to chain the computations by linking 
the inputs and outputs of computer codes, and then to 
execute these computations on remote machines.

Several studies have been conducted at EDF 
based on the \ot component of Salome. 
For example, an uncertainty propagation study (the thermal evaluation 
of the storage of high-level nuclear waste) was making use of a 
computer code where one single run required approximately 10 minutes 
on the 8 cores of a workstation (with shared memory).
Within Salome, the \ot simulation involving 6000 unitary evaluations 
of the function G required 8000 CPU hours on 32 nodes \cite{Barate2013}.

\section{Conclusions}
This educational example has shown a number of questions and problems that can be addressed by UQ methods: uncertainty quantification, central tendency evaluation, excess probability assessment and sensitivity analysis, that can require the use of a metamodel.

Different numerical methods have been used for solving these three classes of problems, leading substantially to the same (or very similar) results. In the industrial practice of UQ, the main issue (which actually motivates the choice of one mathematical method instead of another) is the computational budget, which is actually given by the number of allowed runs of the deterministic model $G$. When the computer code implementing $G$ is computationally expensive, one needs specifically designed mathematical and software tools.

\ot is specially intended to meet these issues : (i) it includes a set of efficient mathematical methods for UQ and (ii) it can be easily connected to any external black box model $G$. Thanks to these two main features, \ot is a software that can address many different physics problems, and thus help to solve industrial problems. From this perspective, the partnership around \ot focuses efforts on the integration of the most efficient and innovative methods required by the industrial applications that takes into account both the need of genericity and of ease to communicate. The main projects for 2015 concern the improvement of the kriging implementation to integrate some very smart methods of optimization. Around this theme some other classical optimization methods will also be generalized or newly implemented.

A growing need for model exploration and analysis of uncertainty problem in industrial applications is to better visualise the information provided by such a volume of data. In this area, specific visualization software, such as Paraview, can provide very efficient and interactive features. Taking the advantage of the integration of \ot in the Salome platform, EDF is working on a better link between the Paraview module in Salome (called ParaVIS) and the uncertainty analysis with \ot : in 2012, functional boxplot (\citep{Hyndman10}) has been implemented. Some recent work around in-situ visualization for uncertainty analysis should also be developed and implemented and so benefit very computationaly expensive model physics that generate an extremely high volume of data. 

Part of this work has been backed by French National Research Agency (ANR) trough the Chorus project (no. ANR-13-MONU-0005-08). We are grateful to the \ot Consortium members. We also thank Regis Lebrun, Mathieu Couplet and Merlin Keller for their help.



\bibliographystyle{spbasic}
\bibliography{bibli}

\begin{thebibliography}{40}
\providecommand{\natexlab}[1]{#1}
\providecommand{\url}[1]{{#1}}
\providecommand{\urlprefix}{URL }
\expandafter\ifx\csname urlstyle\endcsname\relax
  \providecommand{\doi}[1]{DOI~\discretionary{}{}{}#1}\else
  \providecommand{\doi}{DOI~\discretionary{}{}{}\begingroup
  \urlstyle{rm}\Url}\fi
\providecommand{\eprint}[2][]{\url{#2}}

\bibitem[{Airbus et~al(2014)Airbus, EDF, and Phimeca}]{OTDevGuide14}
Airbus, EDF, Phimeca (2014) Developer's guide, {OpenTURNS} 1.4,
  \url{http://openturns.org}

\bibitem[{Au and Beck(2001)}]{au_beck01}
Au S, Beck JL (2001) Estimation of small failure probabilities in high
  dimensions by subset simulation. Probabilistic Engineering Mechanics
  16:263--277

\bibitem[{Barate(2013)}]{Barate2013}
Barate R (2013) Calcul haute performance avec {OpenTURNS}, workshop du {GdR}
  {MASCOT-NUM}, {Quantification} d'incertitude et calcul intensif",
  \url{http://www.gdr-mascotnum.fr/media/openturns-hpc-2013-03-28.pdf}

\bibitem[{Berg(2014)}]{MuParserLib}
Berg I (2014) muparser, \urlprefix\url{http://muparser.beltoforion.de}, fast
  Math Parser Library

\bibitem[{Berger(1985)}]{berger}
Berger J (ed)  (1985) Statistical decision theory and bayesian analysis.
  Springer

\bibitem[{Blatman(2009)}]{Blatman}
Blatman G (2009) Adaptative sparse polynomial chaos expansions for uncertainty
  propagation and sensitivity anaysis. PhD thesis, Clermont University

\bibitem[{Blatman and Sudret(2011)}]{blasud11}
Blatman G, Sudret B (2011) Adaptive sparse polynomial chaos expansion based on
  least angle regression. Journal of Computational Physics 230:2345--2367

\bibitem[{Butucea et~al(2015)Butucea, Delmas, Dutfoy, and
  Fischer}]{Fischer2015}
Butucea C, Delmas J, Dutfoy A, Fischer R (2015) Maximum entropy copula with
  given diagonal section. Journal of Multivariate Analysis, in press

\bibitem[{Ditlevsen and Madsen(1996)}]{Ditlevsen1996}
Ditlevsen O, Madsen H (1996) Structural Reliability Methods. Wiley

\bibitem[{Dutfoy et~al(2009)Dutfoy, Dutka-Malen, Pasanisi, Lebrun, Mangeant,
  Gupta, Pendola, and Yalamas}]{dutdut09}
Dutfoy A, Dutka-Malen I, Pasanisi A, Lebrun R, Mangeant F, Gupta JS, Pendola M,
  Yalamas T (2009) {OpenTURNS, An Open Source initiative to Treat
  Uncertainties, Risks'N Statistics in a structured industrial approach}. In:
  Proceedings of 41\`emes Journ\'ees de Statistique, Bordeaux, France

\bibitem[{Fang et~al(2006)Fang, Li, and Sudjianto}]{fanli06}
Fang KT, Li R, Sudjianto A (2006) Design and modeling for computer experiments.
  Chapman \& Hall/CRC

\bibitem[{gum08(2008)}]{gum2008}
gum08  (2008) JCGM 100-2008 - Evaluation of measurement data - Guide to the
  expression of uncertainty in measurement. JCGM

\bibitem[{Hyndman and Shang(2010)}]{Hyndman10}
Hyndman R, Shang H (2010) Rainbow plots, bagplots, and boxplots for functional
  data. Journal of Computational and Graphical Statistics 19:29--45

\bibitem[{Iooss and Lema\^{\i}tre(2015)}]{ioolem15}
Iooss B, Lema\^{\i}tre P (2015) A review on global sensitivity analysis
  methods. In: Meloni C, Dellino G (eds) Uncertainty management in
  Simulation-Optimization of Complex Systems: Algorithms and Applications,
  Springer

\bibitem[{Kurowicka and Cooke(2006)}]{kurcoo06}
Kurowicka D, Cooke R (2006) Uncertainty analysis with high dimensional
  dependence modelling. Wiley

\bibitem[{Lebrun and Dutfoy(2009{\natexlab{a}})}]{lebdut09t}
Lebrun R, Dutfoy A (2009{\natexlab{a}}) Do rosenblatt and nataf
  isoprobabilistic transformations really differ? Probabilistic Engineering
  Mechanics 24:577--584

\bibitem[{Lebrun and Dutfoy(2009{\natexlab{b}})}]{lebdut09b}
Lebrun R, Dutfoy A (2009{\natexlab{b}}) A generalization of the nataf
  transformation to distributions with elliptical copula. Probabilistic
  Engineering Mechanics 24:172--178

\bibitem[{Lebrun and Dutfoy(2009{\natexlab{c}})}]{lebdut09}
Lebrun R, Dutfoy A (2009{\natexlab{c}}) An innovating analysis of the nataf
  transformation from the viewpoint of copula. Probabilistic Engineering
  Mechanics 24:312--320

\bibitem[{Lebrun and Dutfoy(2009{\natexlab{d}})}]{lebdut09q}
Lebrun R, Dutfoy A (2009{\natexlab{d}}) A practical approach to dependence
  modelling using copulas. Journal of Risk and Reliability 223(04):347--361

\bibitem[{Lebrun and Dutfoy(2014)}]{lebdut14}
Lebrun R, Dutfoy A (2014) Copulas for order statistics with prescribed margins.
  Journal of Multivariate Analysis 128:120--133

\bibitem[{Lemaire(2009)}]{lema09}
Lemaire M (2009) Structural reliability. Wiley

\bibitem[{Liberty(2003)}]{Ev3Lib}
Liberty L (2003) Ev3: A library for symbolic computation in c++ using n-ary
  trees, \urlprefix\url{http://www.lix.polytechnique.fr/~liberti/Ev3.pdf}

\bibitem[{Marin and Robert(2007)}]{marin}
Marin JM, Robert C (eds)  (2007) Bayesian core: a practical approach to
  computational bayesian statistics. Springer

\bibitem[{Marrel et~al(2008)Marrel, Iooss, {Van Dorpe}, and Volkova}]{marioo08}
Marrel A, Iooss B, {Van Dorpe} F, Volkova E (2008) An efficient methodology for
  modeling complex computer codes with {G}aussian processes. Computational
  Statistics and Data Analysis 52:4731--4744

\bibitem[{Munoz-Zuniga et~al(2011)Munoz-Zuniga, Garnier, and Remy}]{mungar11}
Munoz-Zuniga M, Garnier J, Remy E (2011) Adaptive directional stratification
  for controlled estimation of the probability of a rare event. Reliability
  Engineering and System Safety 96:1691--1712

\bibitem[{Nash(2000)}]{nash00}
Nash S (2000) A survey of truncated-newton methods. Journal of Computational
  and Applied Mathematics 124:45--59

\bibitem[{OPEN~CASCADE(2006)}]{ope06}
OPEN~CASCADE S (2006) Salome: The open source integration platform for
  numerical simulation. \urlprefix\url{http://www.salome-platform.org}

\bibitem[{Pasanisi(2014)}]{pas14}
Pasanisi A (2014) Uncertainty analysis and decision-aid: {M}ethodological,
  technical and managerial contributions to engineering and {R\&D} studies.
  Habilitation Thesis of Universit\'e de Technologie de Compi\`egne, France,
  \urlprefix\url{https://tel.archives-ouvertes.fr/tel-01002915}

\bibitem[{Pasanisi and Dutfoy(2012)}]{pasdut12}
Pasanisi A, Dutfoy A (2012) An industrial viewpoint on uncertainty
  quantification in simulation: Stakes, methods, tools, examples. In:
  Dienstfrey A, Boisvert R (eds) Uncertainty quantification in scientific
  computing - 10th IFIP WG 2.5 working conference, WoCoUQ 2011, Boulder, CO,
  USA, August 1-4, 2011, Berlin: Springer, IFIP Advances in Information and
  Communication Technology, vol 377, pp 27--45

\bibitem[{Rasmussen et~al(2006)Rasmussen, Williams, and Dietterich}]{Rasmussen}
Rasmussen C, Williams C, Dietterich T (2006) Gaussian processes for machine
  learning. MIT Press

\bibitem[{Robert and Casella(2004)}]{Robert2004}
Robert CP, Casella G (2004) Monte Carlo statistical methods. Springer

\bibitem[{Rubinstein(1981)}]{Rubinstein81}
Rubinstein R (1981) Simulation and The Monte-Carlo methods. Wiley

\bibitem[{Sacks et~al(1989)Sacks, Welch, Mitchell, and Wynn}]{sacwel89}
Sacks J, Welch W, Mitchell T, Wynn H (1989) Design and analysis of computer
  experiments. Statistical Science 4:409--435

\bibitem[{Saltelli(2002)}]{sal02}
Saltelli A (2002) Making best use of model evaluations to compute sensitivity
  indices. Computer Physics Communication 145:280--297

\bibitem[{Saltelli et~al(1999)Saltelli, Tarantola, and Chan}]{saltar99}
Saltelli A, Tarantola S, Chan K (1999) A quantitative, model-independent method
  for global sensitivity analysis of model output. Technometrics 41:39--56

\bibitem[{Saltelli et~al(2000)Saltelli, Chan, and Scott}]{salcha00}
Saltelli A, Chan K, Scott E (eds)  (2000) Sensitivity analysis. Wiley Series in
  Probability and Statistics, Wiley

\bibitem[{Santner et~al(2003)Santner, Williams, and Notz}]{Santner}
Santner T, Williams B, Notz W (2003) The design and analysis of computer
  experiments. Springer

\bibitem[{Sudret(2008)}]{sud08}
Sudret B (2008) Global sensitivity analysis using polynomial chaos expansion.
  Reliability Engineering and System Safety 93:964--979

\bibitem[{Tarantola(2005)}]{tar05}
Tarantola A (2005) Inverse problem theory and methods for model parameter
  estimation. Society for Industrial and Applied Mathematics, SIAM

\bibitem[{{Top 500 Supercomputer Sites}(2014)}]{ZumbrotaTop500}
{Top 500 Supercomputer Sites} (2014) Zumbrota,
  \urlprefix\url{http://www.top500.org/system/177726}, {B}lue{G}ene/{Q},
  {P}ower {BQC} 16{C} 1.60{GH}z, {C}ustom

\end{thebibliography}

\end{document}